\documentclass[12pt]{JHEP3}
\usepackage{amsmath}
\usepackage{amssymb}
\usepackage{pifont}

\usepackage{epsfig,multicol,bbm}
\usepackage{subfigure}

\title{Hawking-Page Phase Transition of black Dp-branes and R-charged black holes with an IR Cutoff}

\author{Rong-Gen Cai \\
Kavli Institute for Theoretical Physics China (KITPC) at the
Chinese
Academy of Sciences \\
 Institute of Theoretical Physics,
 Chinese Academy of Sciences\\
 P.O.Box 2735, Beijing 100080, China\\
 \email{cairg@itp.ac.cn}}
\author{Li-Ming Cao, Ya-Wen Sun \\
Institute of Theoretical Physics,
Chinese Academy of Sciences\\
P.O.Box 2735, Beijing 100080, China\\
Graduate University of the Chinese Academy of Sciences, \\
Beijing 100049, China\\
\email{caolm@itp.ac.cn; sunyw@itp.ac.cn}}

\preprint{CAS-KITPC/ITP-010}

\abstract{We show that the confinement-deconfinement phase
transition of supersymmetric Yang-Mills theories with 16
supercharges in various dimensions can be realized through the
Hawking-Page phase transition between the near horizon geometries of
black Dp-branes and BPS Dp-branes by removing a small radius region
in the geometry in order to realize a confinement phase, which
generalizes Herzog's discussion for the holographic hard-wall
AdS/QCD model. Removing a small radius region in the gravitational
dual corresponds to introducing an IR cutoff in the dual field
theory. We also discuss the Hawking-Page phase transition between
thermal $AdS_5$, $AdS_4$, $AdS_7$ spaces and R-charged AdS black
holes coming from the spherical reduction of the decoupling limit of
rotating D3-, M2-, and M5- branes in type IIB supergravity and 11
dimensional supergravity in grand canonical ensembles, where the IR
cutoff also plays a crucial role in the existence of the phase
transition.}


\begin{document}


\section{Introduction}
The AdS/CFT
correspondence~\cite{Maldacena:1997re,Gubser:1998bc,Witten:1998qj,mal2}
conjectures that type II$B$ string theory on $AdS_5\times S^5$ is
dual to ${\mathcal{N}}=4$ $SU(N)$ supersymmetric Yang-Mills (SYM)
theory in 3+1 dimensions. At low energies, the string theory can be
approximated by supergravity on $AdS_5$, while the SYM theory is a
conformal field theory on the boundary of $AdS_5$. At finite
temperature, Witten related the Hawking-Page phase transition of
black holes in $AdS_5$ space with the confinement-deconfinement
phase transition of dual SYM~\cite{Witten:1998zw}. On the gravity
side, there are two classical solutions with the same boundary: the
thermal AdS space and the Schwarzschild-AdS black hole which
approaches $AdS_5$ asymptotically. As noted first by Hawking and
Page~\cite{Hawking:1982dh}, a first order phase transition occurs at
some critical temperature, above which an AdS black hole forms. On
the other hand, at a lower temperature, the thermal gas in $AdS_5$
dominates. This Hawking-Page phase transition is identified with the
first order confinement-deconfinement phase transition of dual SYM
theory: at low temperature, the field theory is in a confinement
phase and above a critical temperature it is in a deconfinement
phase.

In Witten's example, the boundary on which the finite temperature
field theory lives is a compact space $S^1\times S^3$. The radius
of the 3 dimensional sphere breaks the conformal symmetry of the
field theory, which makes the phase transition possible. For the
case with a non-compact boundary $S^1\times R^3$, because of the
conformal invariance, no Hawking-Page phase transition exists and
on the SYM side only the deconfinement phase is present even in a
finite temperature
case~\cite{Maldacena:1998im,Rey:1998ik,wilsonloop}. However, the
authors of Ref.~\cite{Polchinski:2001tt} are able to realize
confinement in certain supersymmetric theories by removing a small
radius part of the AdS geometry when the boundary is noncompact.
In the framework of gauge/gravity correspondence, the radial
coordinate on the gravity side corresponds to the energy scale on
the field theory side. Thus the small radius cutoff on the gravity
side implies introducing  an IR cutoff in the field theory. The
so-called hard wall AdS/QCD model has been extensively employed in
discussing various properties of low energy
QCD~\cite{Herzog:2006ra,BoschiFilho:2006pe,Kim:2006ut,Kajantie:2006hv,Kim:2007qk,Kim:2007rt,Ballon
Bayona:2007vp,Cai:2007zw,Cai:2007new,sin}.

Then there is one point to remind here that for supersymmetric
field theories which live on a non-compact space, introducing an
IR cutoff is an effective way to realize a
confinement-deconfinement phase transition, while for those which
live on a flat but at least one dimension compact space, ie.
$S^1\times T^3$ or so, there is a kind of AdS
soliton~\cite{Horowitz:1998ha} which can be used to realize
confinement. Hawking-Page phase transitions can occur between
Ricci flat AdS black holes and AdS solitons both with at least one
dimension compact, see, for
example,~\cite{Surya:2001vj,caigauss,Banerjee:2007by}.

AdS/CFT correspondence was first noticed by Maldacena when studying
the decoupling limit of N coincident D3-branes. In the case of
coincident Dp-branes ($p\neq3$), there are also correspondences of
this kind between certain supergravity solutions and SU(N)
supersymmetric field theories with sixteen supercharges in $p+1$
dimensions~\cite{mal3}. In the decoupling limit, the geometry of
supergravity solutions is no longer $AdS$ and in these cases the
field theories are no longer conformal field theories. Although so,
as in the case of D3-branes, the Hawking-Page phase transition does
not happen when the boundary is noncompact, implying these field
theories are in the deconfinement phase. In this paper we will study
the confinement-deconfinement phase transition of these field
theories by introducing an IR cutoff in the dual supergravity
descriptions, which generalizes Herzog's discussion on the
deconfinement transition of hard wall AdS/QCD
model~\cite{Herzog:2006ra}. In the decoupling limit of rotating
black
 D3-branes, M2-brans, and M5-branes, there  also exist correspondences
between R-charged AdS black holes and R-charged supersymmetric field
theories at finite
temperature~\cite{9810225,caicritical,Harmark:1999xt,Cvetic:1999xp,Cvetic:1999rb,Cvetic:1999ne,CEJM}.
In this paper, we will also study the confinement-deconfinement
phase transition of these R-charged supersymmetric field theories
with an IR cutoff in the dual description. Recently, the author
of~\cite{wenwenyu} studied the phase transition of $AdS$ R-charged
black holes. However the black holes discussed there are R-charged
$AdS$ black holes with spherical horizons; while we study the
R-charged $AdS$ black holes with Ricci flat horizons, which come
from the sphere reduction of the decoupling limit of rotating black
D3-, M2-, M5-branes.

The paper is organized as follows. In the next section, as a
warmup exercise, we will briefly review the Hawking-Page phase
transition for $AdS$ black holes with the boundary $S^1\times
R^3$. Then in section 3, we will study the Hawking-Page phase
transition for the general case of near horizon limit of N
coincident black Dp-branes, whose boundaries are non-compact
$S^1\times R^p$. In section 4, we study the case of the R-charged
$AdS_5$, $AdS_4$, and $AdS_7$ black holes, respectively. Sec.~5 is
devoted to conclusions.
\section{Hawking-Page phase transition for Ricci flat black holes with an IR cutoff}

In this section we review the deconfinement transition of hard-wall
AdS/QCD  through the Hawking-Page phase transition between thermal
$AdS_5$ and an $AdS_5$ black hole with a non-compact boundary
$S^1\times R^3$. For more details, see ~\cite{Herzog:2006ra}. In
order to study the phase transition of the boundary CFT using the
gravity description, we should first find the classical solutions of
AdS supegravity with the same asymptotic boundary $S^1\times R^3$,
and then compare the Euclidean actions of these classical solutions
to see if there is a phase transition. However, as we know, the
actions always diverge due to the infinite space. There are two ways
to get a finite result: one is to add surface counterterms to the
action and the other is the so-called background subtraction method
where a suitable reference background is chosen so that the solution
under study can be asymptotically embedded into this background.
Here we use the background subtraction method as it is more suitable
to our purpose to calculate the difference of two Euclidean actions
in this paper.

In the Euclidean sector, the action of 5-dimensional gravity with
a cosmological constant can be written as
\begin{equation}
\label{action1} I=-\frac{1}{16\pi G}\int d^5x\sqrt{g}(R-2\Lambda),
\end{equation}
where $\Lambda$ is the cosmological constant which can be related to
the radius scale $l$ of AdS space by $ \Lambda=-6/l^2 $. According
to the action (\ref{action1}), there are two solutions with the same
asymptotic boundary $S^1\times R^3$, i.e. thermal AdS space and the
AdS black hole solution (in Lorentz sector):
\begin{eqnarray}
\label{2eq4} &&
ds_{AdS}^2=\frac{U^2}{l^2}\left(-dt^2+dx_1^2+dx_2^2+dx_3^2\right)+\frac{l^2}{U^2}dU^2,
\\
&&
ds_{BH}^2=\frac{U^2}{l^2}\left[-\left(1-\frac{U_H^4}{U^4}\right)dt^2+dx_1^2+dx_2^2
+dx_3^2\right]+\frac{l^2}{U^2}\left(1-\frac{U_H^4}{U^4}\right)^{-1}dU^2,
\end{eqnarray}
where $U_H$ corresponds to the horizon of the black hole. After a
Euclidean continuation $t=i\tau$ the two solutions become
\begin{eqnarray}
&&
ds_{AdS}^2=\frac{U^2}{l^2}\left(d\tau^2+dx_1^2+dx_2^2+dx_3^2\right)+\frac{l^2}{U^2}dU^2,
\\
&&
ds_{BH}^2=\frac{U^2}{l^2}\left[\left(1-\frac{U_H^4}{U^4}\right)d\tau^2+dx_1^2+dx_2^2
+dx_3^2\right]+\frac{l^2}{U^2}\left(1-\frac{U_H^4}{U^4}\right)^{-1}dU^2.
\end{eqnarray}
To eliminate the conical singularity, the $\tau$ in the AdS black
hole solution should get a period
\begin{equation}
\label{2eq6}
 \beta =\frac{\pi l^2}{U_H},
\end{equation}
while the period of $\tau$ for the thermal $AdS$ could be
arbitrary.  This period (\ref{2eq6}) is just the inverse of the
temperature of the AdS black hole. To see whether there is a phase
transition between the AdS black hole and thermal AdS space,  we
should calculate the difference of the Euclidean actions for these
two solutions. The Euclidean actions of the AdS black hole and the
thermal AdS are
\begin{eqnarray}
&& I_{BH}=\frac{8}{16\pi Gl^5}\int_{U_H}^{U_{uv}} d^5xU^3,
\\
&& I_{AdS}=\frac{8}{16\pi Gl^5}\int_{0}^{U_{uv}}d^5xU^3,
\end{eqnarray}
respectively. Here to get the actions of both solutions, we have
introduced a finite UV boundary at $U=U_{uv}$. At the end of
calculations, the limit $U_{uv}\rightarrow\infty$ will be taken. At
the boundary,  the temperatures for both solutions should be the
same. This means that we have the following relation for the two
temperatures
\begin{equation}
\beta_{AdS}=\beta \sqrt{1-\frac{U_H^4}{U_{uv}^4}}.
\end{equation}
It turns out that the difference of the two actions is
\begin{equation}
\Delta
I=\lim_{U_{uv}\rightarrow\infty}\left(I_{BH}-I_{AdS}\right)=-\frac{V(\vec{x})U_H^4\beta}{16\pi
Gl^5}<0,
\end{equation}
where $V(\vec{x})$ denotes the volume of the three flat dimensions
$x_1$, $x_2$ and $x_3$. This negative action difference means that
the black hole always dominates and confirms that the dual field
theory is in the deconfinement phase. Now we introduce an IR cutoff
$U_{IR}$ in the coordinates (\ref{2eq4}), where the IR cutoff
$U_{IR}$ is equivalent to an IR cutoff (mass gap) in the dual field
theory, then the integral in the action of thermal AdS should start
from $U_{IR}$ and the integral of the black hole geometry should
start from
$U_{max}=$max$[U_{IR},U_{H}]$~\cite{Herzog:2006ra,Cai:2007zw}. Now
the Euclidean actions of the two solutions are
\begin{equation}
I_{BH}=\frac{V(\vec{x})\beta}{16\pi
Gl^5}\left(U_{uv}^4-U_{max}^4\right),
\end{equation}
and
\begin{equation}
I_{AdS}=\frac{V(\vec{x})\beta_{AdS}}{16\pi
Gl^5}\left(U_{uv}^4-U_{IR}^4\right),
\end{equation}
respectively.
 Thus one has the action difference
\begin{equation}
\Delta I=\lim_{U_{uv}\rightarrow\infty}=\frac{V(\vec{x})\beta}{16\pi
Gl^5}\left(\frac{1}{2}U_H^4-U^4_{max}+U_{IR}^4\right).
\end{equation}
The action difference obviously depends on the IR cutoff and
$U_{max}$. When the temperature is very low, $U_H$ is small compared
to $U_{IR}$, one then has  $U_{max}=U_{IR}$, and
\begin{equation}
\label{2eq14}
 \Delta I=\frac{V(\vec{x})\beta}{16\pi
Gl^5}\frac{1}{2}U_H^4>0.
\end{equation}
On the other hand, when the temperature gets higher, $U_H$ will
become larger than $U_{IR}$, one takes $U_{max}=U_{H}$, and has
\begin{equation}
\label{2eq15}
 \Delta I=\frac{V(\vec{x})\beta}{16\pi
Gl^5}\left(U_{IR}^4-\frac{1}{2}U_H^4\right).
\end{equation}
Eq.~(\ref{2eq14}) tells us that in the low temperature phase, where
$U_{IR} > U_{H}$, the thermal gas in $AdS$ dominates and there is no
Hawking-Page transition; and it implies that the dual field theory
is in the confinement phase.  However, when $U_{IR}< U_H$,  the
action difference (\ref{2eq15}) can change its sign from positive to
negative at a critical temperature where
$U_{IR}^4=\frac{1}{2}U_H^4$. The critical temperature is
\begin{equation}
\label{2eq16}
 \beta_{crit}=\frac{\pi l^2}{2^{\frac{1}{4}}U_{IR}}.
\end{equation}
The Hawking-Page transition indicates that when $T>1/\beta_{crit}$,
the AdS black hole dominates, while the thermal AdS space dominates
when $T<1/\beta_{crit}$. In the dual field theory side, the field
theory is in the deconfinement phase at $T> 1/\beta_{crit}$, while
it is in the confinement phase at $T<1/\beta_{crit}$. When the
temperature $T$ crosses the critical temperature $1/\beta_{crit}$,
the deconfinement phase transition happens.  The IR cutoff $U_{IR}$
can be related to the mass of the lightest meson in the holographic
AdS/QCD model~\cite{Herzog:2006ra}.  As a result, we see that an IR
cutoff can realize the Hawking-Page transition for Ricci flat AdS
black holes when the boundary is non-compact. It is easy to
understand the occurrence of the transition because an IR cutoff
breaks the conformal symmetry for the dual field theory.  Finally,
we mention here that usually the Gibbons-Hawking surface term should
be included in calculating the Euclidean action of black holes.
However, for asymptotically AdS spacetimes it turns out that the
surface term will not make a contribution to the action
difference~\cite{Hawking:1982dh}, which will be clearly seen in the
next section.

\section{Hawking-Page phase transition in black Dp-branes with IR cutoff}

Like the D3-branes, the decoupling limit of Dp-branes in type II
supergravity has also field theory description; they are
supersymmetric Yang-Mills theories with 16 supercharges in $p+1$
dimensions~\cite{mal3}.  In this section, we will generalize the
discussions in Sec.~2 to the cases of those finite temperature
non-conformal field theories defined on the boundary $S^1\times
R^p$ by studying the decoupling limit of Dp-branes. Generally to
get a well-defined decoupling limit,  $p$ should be limited the
range $0\le p\le 4$.

To see whether there will be a phase transition for dual field
theories at finite temperature, we will first get the two
classical Euclidean solutions with the same asymptotic boundary
$S^1\times R^p$, and then compare the Euclidean actions of the two
solutions in both cases with and without an IR cutoff. The two
classical solutions can be obtained by taking the decoupling
limits of black Dp-branes and BPS Dp-branes.
\subsection{The decoupling limit of black Dp-branes}
Black Dp-branes are non-BPS solutions of ten dimensional Type
$\amalg$ supergravities. The bulk action of the supergravity is
\begin{equation}
S_{str}=-\frac{1}{16\pi G_{10}} \int d^{10}x \sqrt{-g}\left[e^{-2
\phi}\left(R+4(\partial \phi)^2\right)-\frac{1}{2d!}F^2_{d}\right]
\end{equation}
in string frame, and
\begin{equation}
\label{action2} S_{Ein}=-\frac{1}{16\pi G_{10}}\int d^{10}x
\sqrt{-{g}}\left[R-\frac{1}{2}(\partial
\phi)^2-\frac{e^{-\alpha(d) \phi}}{2d!}F^2_{d}\right]
\end{equation}
in Einstein frame, where $ d=p+2 $ in the case of the electric
brane and $ d=8-p $ of the magnetic brane. $\alpha(d)$ depends on
the value of $d$: $ \alpha(d)=\frac{d-5}{2}. $ Since $F_{8-p}$ and
$F_{p+2}$ both do not change under the frame transformation while
the metric changes, the duality relation changes from $
F_{8-p}=*F_{p+2} $ in string frame to $
F_{8-p}=e^{-\alpha(p+2)\phi}*F_{p+2} $ in Einstein frame.

The solution for $N$ coincident black Dp-branes is
\begin{eqnarray}
&& ds_{s}^2=H^{-\frac{1}{2}}(r)\left (-f(r)dt^2+
\sum_{i=1}^{p}{(dx^i)^2}\right )+H^{\frac{1}{2}}(r)\left
(f^{-1}(r)dr^2+r^2d\Omega^2_{8-p}\right ),
\\ && e^\phi=g_sH^{\frac{3-p}{4}}, \\
&& A_{t1\cdots
p}=g^{-1}_s\left(1-H^{-1}\right)\coth\beta.
\end{eqnarray}
in string frame, where
\begin{eqnarray}
&&
H(r)=1+\xi\frac{c_pg_sN\alpha'^{\frac{7-p}{2}}}{r^{7-p}}=1+\xi~\frac{r^{7-p}_p}{r^{7-p}}
=1+\sinh^2\beta\frac{r^{7-p}_H}{r^{7-p}}, \\
&& c_p\equiv (2\sqrt
\pi)^{5-p}\Gamma\left[\frac{1}{2}(7-p)\right], \\
&&  \xi=\tanh \beta=\sqrt{1+\left(
\frac{r^{7-p}_H}{2r^{7-p}_p}\right)^2}-\frac{r^{7-p}_H}{2r^{7-p}_p},
\\
&& f(r)=1-\frac{r^{7-p}_H}{r^{7-p}}.
\end{eqnarray}
To get BPS Dp-branes, one can simply set $r_H=0$ and then $f=1$.

We can get the decoupling limit of this solution keeping the
energies fixed by changing
 the parameters to~\cite{mal3}
\begin{eqnarray}
&& N g^2_{YM}=N (2\pi)^{p-2}g_s
\alpha'^{\frac{p-3}{2}}=\mathrm{fixed},
\\
&& U=\frac{r}{\alpha'}=\mathrm{fixed},
\end{eqnarray}
and setting
\begin{equation}
\alpha'\rightarrow0,\quad U_H=\frac{r_H}{\alpha'},\quad
d_p=c_p(2\pi)^{2-p}.
\end{equation}
In this decoupling limit, the solution in Einstein frame becomes
\begin{eqnarray}
&& ds^2_{Ein}=\alpha'^{\frac{7-p}{4}}\Bigg{\{}
\frac{U^{\frac{(7-p)^2}{8}}}{(g^2_{YM}d_pN)^{\frac{7-p}{8}}}\left[-\left(1-\frac{U^{7-p}_H}{U^{7-p}}\right)dt^2
+d\vec{x}^2\right]\nonumber
\\
&&~~~~~~~~~+~\frac{(g^2_{YM}d_pN)^{\frac{p+1}{8}}}{U^{\frac{(p+1)(7-p)}{8}}}\left[\frac
{dU^2}{1-\frac{U^{7-p}_H}{U^{7-p}}}+U^2d\Omega^2_{8-p}\right]\Bigg{\}},
\\
&& e^{\phi}=\alpha'^{\frac{p-3}{2}}\left(
\frac{g_{YM}^2d_pN}{U^{7-p}}\right), \\
&& F_{U01 \cdots
p}=\alpha'^{\frac{p+1}{2}}\frac{(p-7)(2\pi)^{p-2}U^{6-p}}{d_pNg_{YM}^4}.
\end{eqnarray}
This solution is just the gravitational dual of SYM theory with 16
supercharges in $p+1$ dimensions. When $p=3$, the solution turns out
to be $AdS_5\times S^5$, the dual theory is the ${\cal N}=4$ SYM
theory with 32 charges. In that case, the theory is a conformal one.
This case is just discussed in the previous section. Now we study
the general cases without the conformal symmetry.

The Euclidean sector of the above solution can be obtained by
setting $t=i\tau$
\begin{eqnarray}
ds^2_{Euc}&=&\alpha'^{\frac{7-p}{4}}\Bigg{\{}
\frac{U^{\frac{(7-p)^2}{8}}}{(g^2_{YM}d_pN)^{\frac{7-p}{8}}}\left[\left(1
 -\frac{U^{7-p}_H}{U^{7-p}}\right)d\tau^2
+d\vec{x}^2\right]\nonumber
\\
&&~~~~~~~~+
\frac{(g^2_{YM}d_pN)^{\frac{p+1}{8}}}{U^{\frac{(p+1)(7-p)}{8}}}\left[\frac
{dU^2}{1-\frac{U^{7-p}_H}{U^{7-p}}}+U^2d\Omega^2_{8-p}\right]\Bigg{\}}.
\end{eqnarray}
The Euclidean time $\tau$ has a period
\begin{equation}
\label{3eq17}
 \beta=\left
.\frac{4\pi}{\sqrt{\partial_Ug_{\tau\tau}\partial_Ug^{-1}_{UU}}}\right|_{U=U_H}
=\frac{4\pi g_{YM} \sqrt{d_pN}}{(7-p)U^{\frac{5-p}{2}}_H},
\end{equation}
in order to remove the conical singularity. This  is nothing but
 the inverse Hawking temperature of the black Dp-branes in the decoupling limit.

\subsection{Phase transition with an IR cutoff}

To see whether there is a phase transition between the decoupling
limits of black Dp-branes and BPS Dp-branes, or say, deconfinement
transition of those SYM theories at finite temperature,  we first
calculate the on-shell action of those black Dp-branes.  To avoid
the complex surface term in the action for Dp-branes with electric
charge, we consider black Dp-branes with magnetic charge. The
on-shell Euclidean action can be written out using the equation of
motion
\begin{equation}
I =\frac{7-p}{4}\frac{g_s^2}{16\pi G_{10}}\int
d^{10}x\sqrt{g}\frac{e^{-\alpha(8-p) \phi } {F}^2_{8-p}}{2(8-p)!}.
\end{equation}
Note the relation
\begin{equation}
\frac{e^{-\alpha(8-p) \phi
}F^2_{8-p}}{2(8-p)!}=\frac{e^{-\alpha(p+2) \phi
}F^2_{p+2}}{2(p+2)!},
\end{equation}
and one has then the Euclidean action
\begin{equation}
I_{bulk}=\alpha'^{7-p}\frac{(7-p)^3}{8}\frac{V(\vec{x})V(\Omega_{8-p})\beta}{16\pi
G_{10}} \int U^{6-p}dU,
\end{equation}
where $V(\vec{x})$ is the volume of the $p$ spatial dimensions and
$V(\Omega_{8-p})$ is the volume of unit $8-p$ sphere. The factor
of $\alpha'^{7-p}$ can be absorbed into the redefinition of the
Newton constant $G_{10}=8\pi^6g_s^2\alpha'^4=\alpha'^{7-p}8\pi^6
g_{YM}^4/(2\pi)^{2p-4} \equiv \alpha'^{7-p}G_{10}'$ in the
decoupling limit.

We first calculate the difference of the bulk actions of the
decoupling limits of the black Dp-branes and BPS Dp-branes. To
regularize the actions, we introduce a UV boundary $U_{uv}$ for both
solutions, where the local temperatures are the same for both
solutions. Here we use $I_{bl}$ as the Euclidean action of the
decoupling limit of the black Dp-branes and $I_{ba}$ as the
Euclidean action of the decoupling limit of the BPS Dp-branes. Thus
we have
\begin{equation}
I_{bulk}^{bl}=\frac{(7-p)^3}{8}\frac{V(\vec{x})V(\Omega_{8-p})\beta}{16\pi
G_{10}'} \int_{U_H}^{U_{uv}} U^{6-p}dU,
\end{equation}
and
\begin{equation}
I_{bulk}^{ba}=\frac{(7-p)^3}{8}\frac{V(\vec{x})V(\Omega_{8-p})\beta'}{16\pi
G_{10}'} \int_{0}^{U_{uv}} U^{6-p}dU,
\end{equation}
where
\begin{equation}
\beta'=\beta\sqrt{1-\frac{U_H^{7-p}}{U_{uv}^{7-p}}}.
\end{equation}
The difference of these two actions is
\begin{eqnarray}
\Delta I_{bulk} &=&\lim_{U_{uv}\rightarrow \infty}
(I_{bl}-I_{ba})\nonumber \\
&=&\lim_{U_{uv}\rightarrow
\infty}\frac{(7-p)^2}{8}\frac{V_pV(\Omega_{8-p})\beta}{16\pi
G_{10}'}\left[U_{uv}^{7-p}\left(1-\sqrt{1-\frac{U_H^{7-p}}{U_{uv}^{7-p}}}\right)-U_H^{7-p}\right]\nonumber\\
&=&\frac{(7-p)^2}{8}\frac{V_pV(\Omega_{8-p})\beta}{16\pi
G_{10}'}\left(-\frac{1}{2}\right)U_H^{7-p}<0.
\end{eqnarray}
Besides the bulk part, we should also consider the contribution of
the Gibbons-Hawking surface term
\begin{equation}
I_{GB}=-\frac{1}{8\pi G_{10}}\int_{\partial M} d^9x\sqrt{h}K,
\end{equation}
where $h$ is the determinant of the reduced metric on the UV
boundary $\partial M$ and $K$ is the extrinsic curvature of the
reduced metric
\begin{equation}
K=\bigtriangledown_\mu n^{\mu}.
\end{equation}
The surface terms for both solutions are
\begin{eqnarray}
I_{GB}^{bl}&=&-\frac{V(\vec{x})V(\Omega_{8-p})\beta}{16\pi
G_{10}'}\Bigg{[}\left(16-2p-\frac{(7-p)(p+1)}{8}\right)U_{uv}^{7-p}\nonumber\\
&&-\left(9-p-\frac{(7-p)(p+1)}{8}\right)U^{7-p}_H\Bigg{]},
\end{eqnarray}
and
\begin{equation}
I_{GB}^{ba}=-\frac{V(\vec{x})V(\Omega_{8-p})\beta'}{16\pi
G_{10}'}\left[\left(16-2p-\frac{(7-p)(p+1)}{8}\right)U_{uv}^{7-p}\right],
\end{equation}
respectively. Then the difference of the two surface terms is
\begin{equation}
\label{GHpbrane} \Delta
I_{GB}=\frac{V(\vec{x})V(\Omega_{8-p})\beta}{16\pi
G_{10}'}\left(\frac{p-3}{4}\right)^2U_H^{7-p}.
\end{equation}
When $p=3$,  this term vanishes. This confirms that for AdS black
holes, the Gibbons-Hawking surface term has no contribution to the
Euclidean action difference stated in the previous section.
Finally we get the total Euclidean action difference for those two
solutions
\begin{equation}
\Delta I=\Delta I_{bulk}+\Delta
I_{GB}=-\frac{V(\vec{x})V(\Omega_{8-p})\beta}{16\pi
G_{10}'}\left(\frac{5-p}{2}\right)U_H^{7-p}.
\end{equation}
Here we should note that we choose the total action to be the sum of
the bulk term and the Gibbons-Hawking boundary term. There are no
other surface counterterms like boundary cosmological counterterms
used in~\cite{cai9912013} needed here because we are using the
background subtraction method. And the result we get here is the
same as in~\cite{cai9912013} where they also calculated the
Euclidean action of the near horizon geometry of black Dp-branes
(compactified on transverse spheres) but using the counterterm
approach.

Then from the equation above we can see that this action difference
is always negative and hence no Hawking-Page phase transition
happens. This means that the near horizon geometries of black
Dp-branes dominate all the times, and the dual field theories are
always in the deconfinement phase. Now as in the hard-wall AdS/QCD
model, we introduce an IR cutoff to realize a confinement phase.
Correspondingly, we introduce an IR cutoff $U_{IR}$ in the dual
gravitational description by removing the part with $U<U_{IR}$ of
geometry. With the IR cutoff, the integral in the action starts from
$U_{IR}$ in the case of the near horizon limit of Dp-branes and
$U_{max}=$max$[U_{IR},U_{H}]$ in the case of the near horizon limit
of black Dp-branes. In this case, the difference of the actions
becomes
\begin{eqnarray}
\Delta I_{bulk}&=&\lim_{U_{uv}\rightarrow \infty}
(I^{bl}_{bulk}-I^{ba}_{bulk})\nonumber \\
&=&\frac{(7-p)^2}{8}\frac{V(\vec{x})V(\Omega_{8-p})\beta}{16\pi
G_{10}'}\left(\frac{1}{2}U_H^{7-p}-U_{max}^{7-p}+U_{IR}^{7-p}\right),
\end{eqnarray}
while the part from the Gibbons-Hawking surface term keeps
unchanged, still has the form~(\ref{GHpbrane}). Thus, we have
\begin{eqnarray}
\Delta I&=&\Delta I_{bulk}+\Delta I_{GB}\nonumber
\\
&=&\frac{V(\vec{x})V(\Omega_{8-p})\beta}{16\pi
G_{10}'}\Bigg{(}\frac{p^2-10p+29}{8}
U_H^{7-p}-\frac{(7-p)^2}{8}(U_{max}^{7-p}-U_{IR}^{7-p})\Bigg{)}.
\end{eqnarray}
 When $U_H<U_{IR}$, one has $U_{max}=U_{IR}$, and
\begin{equation}
\Delta I=\frac{V(\vec{x})V(\Omega_{8-p})\beta}{16\pi
G_{10}'}\left[\left(\frac{p^2-10p+29}{8}\right)U_H^{7-p}\right]>0.
\end{equation}
On the other hand, when $U_H>U_{IR}$, we have $U_{max}=U_{H}$, and
\begin{equation}
\Delta I=\frac{V(\vec{x})V(\Omega_{8-p})\beta}{16\pi
G_{10}'}\left(-\frac{5-p}{2}U_H^{7-p}+\frac{(7-p)^2}{8}U_{IR}^{7-p}\right).
\end{equation}
We see that the action can change its sign and the Hawking-Page
phase transition happens when $U_H^{7-p}$ $=$
$\frac{(7-p)^2}{4(5-p)}U_{IR}^{7-p}$. The corresponding  critical
temperature is
\begin{equation}
\beta_{crit}=\frac{4\pi g_{YM}
\sqrt{d_pN}}{(7-p){\left(\frac{(7-p)^2}{4(5-p)}U_{IR}^{7-p}\right)}^{\frac{5-p}{2}}}.
\end{equation}
Because the temperature of the black Dp-branes is proportional to
$U_H^{(5-p)/2}$ (see (\ref{3eq17})), it is easy to see that at low
temperature less than $1/\beta_{crit}$, the decoupling limit of
Dp-branes with IR cutoff dominates, which corresponds to the
confinement phase of dual SYM theories; and at high temperature
above the critical temperature, the decoupling limit of black
Dp-branes dominates, which corresponds to the deconfinement phase
of the dual SYM theories.  In addition, we mention again that here
$p$ is in the range $0 \le p\le 4$.

Thus we have shown that as in the case of the hard-wall AdS/QCD
model, one also can realize the deconfinement transition for $p+1$
dimensional SYM theories residing on non-compact manifold $S^1
\times R^p$ through the first order Hawking-Page phase transition
between the decoupling limits of black Dp-branes and BPS Dp-branes
by introducing an IR cutoff.

\section{Hawking-page phase transition of R-charged $AdS_4$, $AdS_5$, and $AdS_7$
black holes with an IR cutoff}
The decoupling limit of the solution of N coincident rotating black
D3-branes of the ten dimensional type II$B$ supergravity action can
be reduced to 5 dimensions through $S^5$ dimensional reduction,
resulting in a 5 dimensional charged AdS black
hole~\cite{Cvetic:1999xp,Cvetic:1999rb,Cvetic:1999ne,CEJM}.
According to AdS/CFT correspondence, these charged AdS black holes
in five dimensions are dual to R-charged SYM theory living on the
boundary. Also the decoupling limits of the solutions of N
coincident rotating black M2 and M5-branes of the 11 dimensional
supergravity can be reduced to charged $AdS_4$ and $AdS_7$ black
holes through $S^7$ and $S^4$ reductions
respectively~\cite{Cvetic:1999xp}. These R-charged AdS black holes
are black holes with Ricci flat horizon.  In this section we discuss
the Hawking-Page phase transitions of those Ricci flat AdS black
holes in grand canonical ensembles. Note that the Hawking-Page phase
transition in those R-charged AdS black holes with spherical horizon
has been discussed in \cite{Cvetic:1999ne,CEJM}, while it has been
studied for the case with hyperbolic horizon in~\cite{CW}.

\subsection{R-charged $AdS_5$ black holes}
In the case of rotating D3-branes, there are six spatial
dimensions transverse to the branes, so there can be at most 3
angular momenta. Thus after dimensional reduction on $S^5$, there
can be three charges, parameterized by $q_1$, $q_2$ and $q_3$
respectively. The action after spherical reduction
becomes~\cite{Cvetic:1999xp}
\begin{equation}
I=-\frac{1}{16\pi G_5}\int
d^5x\sqrt{-g}\left(R-\frac{1}{2}(\partial\vec{\varphi})^2
-\frac{1}{4}\sum_iX_i^2(F_i)^2+\frac{4}{l^2}\sum_iX_i^{-1}\right),
\end{equation}
where
\begin{equation}
 X_i=e^{-\frac{1}{2}\vec{a}_i\cdot\vec{\varphi}}
\end{equation}
with dilation vectors
\begin{equation}
 \vec{a}_1=\left(\frac{2}{\sqrt{6}},\sqrt{2}\right),\quad
 \vec{a}_2=\left(\frac{2}{\sqrt{6}},-\sqrt{2}\right),\quad \vec{a}_3=\left(-\frac{4}{\sqrt{6}},0\right).
\end{equation}
This is just the action of a $U(1)^3$ truncation of the
${\mathcal{N}}=8$, $SO(6)$ gauged supergravity. The solution after
reduction is a black hole solution of this action with three charges
under the $U(1)^3$ and two scalar fields. This solution is
\begin{equation}
ds^2=-
\left({\mathcal{H}}_1{\mathcal{H}}_2{\mathcal{H}}_3\right)^{-\frac{2}{3}}fdt^2
+\left({\mathcal{H}}_1{\mathcal{H}}_2{\mathcal{H}}_3\right)^{\frac{1}{3}}
\left(f^{-1}dr^2+r^2(dx^2_1+dx^2_2+dx^2_3)\right)\, ,
\end{equation}
\begin{equation}
X_i={{\mathcal{H}}_i}^{-1}({\mathcal{H}}_1{\mathcal{H}}_2{\mathcal{H}}_3)^{\frac{1}{3}},\quad
A_t^i=\frac{\sqrt{\mu}\left(1-{{\mathcal{H}}_i}^{-1}\right)}{q_i},
\end{equation}
where
 \begin{equation}
f=\frac{r^2}{l^2}{\mathcal{H}}_1{\mathcal{H}}_2{\mathcal{H}}_3-\frac{\mu}{r^2},\quad
{\mathcal{H}}_i=1+\frac{q_i^2}{r^2},\quad i=1,2,3
\end{equation}
and $\mu$ is the mass parameter of the AdS black hole.

The black hole has the Hawking temperature $1/\beta$,
\begin{equation}
\beta=\left.\left(\frac{4\pi}{({\mathcal{H}}_1{\mathcal{H}}_2{\mathcal{H}}_3)^{-\frac{1}{2}}
\partial_rf}\right)\right|_{r=r_0},
\end{equation}
where $r_0$ corresponds to the black hole horizon, i.e., the largest
real root of $f(r)=0$,
\begin{equation}
\mu
l^2=r_0^4{\mathcal{H}}_1(r_0){\mathcal{H}}_2(r_0){\mathcal{H}}_3(r_0).
\end{equation}
And the Euclidean action becomes
\begin{equation}
I_{E}=-\frac{1}{16\pi G_5}\int
d^5x\sqrt{g}\Bigg{(}R-\frac{1}{2}(\partial\vec{\varphi})^2
-\frac{1}{4}\sum_iX_i^2(F_i)^2+\frac{4}{l^2}\sum_iX_i^{-1}\Bigg{)}.
\end{equation}

\subsubsection{Euclidean action for $AdS_5$ R-charged black holes}
Before discussing the phase transition, we should state that we work
in the grand canonical ensemble where the chemical potentials of the
ensemble are fixed to certain values. The choice of ensemble is
crucial because in grand canonical ensembles the Euclidean action
can be just identified with the Gibbs free energy, while for
canonical ensembles the Helmholtz free energy should be given by the
Legendre transform of the Euclidean action. Thus here we just need
to calculate the difference of Euclidean actions as before. Now the
background we choose is still the pure thermal $AdS_5$ space with
zero valued charges but constant and maybe nonzero chemical
potentials. Then to discuss the Hawking-Page phase transition
associated with the $AdS_5$ R-charged black holes, we first
calculate the on-shell action by using the Einstein equation. The
Euclidean action for this solution is
\begin{equation}
I_{E}=\frac{4}{3}\frac{V(\vec{x})}{16\pi G_5l^2}\int d\tau
drr\left(6r^2+2(q_1^2+q_2^2+q_3^2)-\sum_{i}\frac{\mu
l^2q_i^2}{(r^2+q_i^2)^{2}}\right).
\end{equation}
We will study the phase transition in the grand canonical
ensemble, where the electric potentials are fixed. We can choose a
certain gauge here to make
\begin{equation}
A_i=\frac{\sqrt{\mu}\left(1-{\mathcal{H}}_i^{-1}\right)}{q_i}+\Phi_i=0
\end{equation}
at the horizon $r=r_0$. The gauge invariant chemical potential
between the black hole horizon and infinity is $\Phi_i$, since
only this quantity enters into the action and other physical
quantities.

To calculate the Euclidean action difference, we have to select an
appropriate background. For the case of R-charged black holes, it is
natural to select the pure AdS space-time with constant chemical
potentials $\Phi_i$, since this background is still the solution of
equations of motion. Next we have to fix the period of Euclidean
time of the pure AdS space-time. This can be done by equating the
induced metric of the pure AdS space-time on the hypersurface
$r=\mathrm{constant}$ with the one of black hole~\cite{chen:beta}.
This means we have
\begin{equation}
\int d\tau d^{3}x\sqrt{h} =\int d\tau' d^3x\sqrt{h'},
\end{equation}
where $h$ and $h'$ are the determinants of the induced metrics of
black hole and the pure AdS space-time. Thus we find
\begin{equation}
\label{period} \beta'=\beta \frac{\int d^{3}x\sqrt{h} }{\int
d^3x\sqrt{h'}},
\end{equation}
where the integration is taken on the $r=r_{uv}$ hypersurface (an
UV boundary). For the 3-charged black hole, we have
\begin{equation}
\label{betabetaprime1} \beta'=\beta \left.
\sqrt{\frac{l^2({\mathcal{H}}_1{\mathcal{H}}_2{\mathcal{H}}_3)^{\frac{1}{3}}f}{r^2}}\right|_{r_{uv}}.
\end{equation}
For convenience we write
\begin{equation}
{\mathcal{H}}_1{\mathcal{H}}_2{\mathcal{H}}_3=\left(1+\frac{q_1^2}{r^2}\right)\left(1+\frac{q_2^2}{r^2}\right)
\left(1+\frac{q_3^2}{r^2}\right)=
1+\frac{A}{r^2}+\frac{B}{r^4}+\frac{C}{r^6},
\end{equation}
with $A, B,C$ defined as follows
\begin{equation}
A=q_1^2+q_2^2+q_3^2,\quad B=q_1^2q_2^2+q_1^2q_3^2+q_2^2q_3^2,\quad
C=q_1^2q_2^2q_3^2.
\end{equation}

Let us first consider the case without an IR cutoff. In this case,
the Euclidean action of the black hole solution is
\begin{equation}
 I^{bl}_{bulk}=\frac{2V(\vec{x})\beta}{16\pi
G_5l^2}\Bigg{[}(r_{uv}^4-r_0^4)+\frac{2A}{3}(r_{uv}^2-r_0^2)+\frac{\mu
l^2}{3}
\sum_i\left(\frac{q_i^2}{r_{uv}^2+q_i^2}-\frac{q_i^2}{r_0^2+q_i^2}\right)\Bigg{]},
\end{equation}
 while for the pure
AdS background, we have
\begin{equation}
I^{ba}_{bulk}
=\frac{2V(\vec{x})\beta'}{16\pi G_5l^2}\left(r_{uv}^4\right).
\end{equation}
Here we should note that just as stated in the beginning of this
subsection, we are working in grand canonical ensemble and the
background thermal $AdS_5$ spacetime has nonzero constant chemical
potentials as in reference~\cite{CEJM}, so the Euclidean action
which only involves the gauge field strength but not the gauge
potential is unaffected by the nonzero fixed potentials.

 The action difference is
\begin{equation}
\Delta I_{bulk}=\frac{V(\vec{x})\beta}{16\pi
G_5l^2}\left(-r_0^4-Ar_0^2-\frac{C}{r_0^2}+\frac{1}{9}(2A^2-15B)\right).
\end{equation}
The contribution of the Gibbons-Hawking surface term for the black
hole solution is
\begin{eqnarray}
I_{GH}^{bl}&=&-\frac{1}{8\pi G_5}\int_{\partial M} d\tau d^3x
\sqrt{h}K  \nonumber \\
&=&\frac{V(\vec{x})\beta}{8\pi G_5
l^2}\left(4r_{uv}^4+\frac{8}{3}Ar_{uv}^2+\left(\frac{4}{3}B-2\mu
l^2\right)+\mathcal{O}\left(\frac{1}{r_{uv}}\right) \right).
\end{eqnarray}
 For the pure AdS background it is
\begin{equation}
I_{GH}^{ba}=-\frac{1}{8\pi G_5}\int_{\partial M} d\tau d^3x
\sqrt{h}K=\frac{V(\vec{x})\beta'}{8\pi G_5
l^2}\left(4r_{uv}^4\right).
\end{equation}
As a result, the part of action difference from the
Gibbons-Hawking surface term is
\begin{equation}
\Delta I_{GH} =\frac{V(\vec{x})\beta}{8\pi G_5
l^2}\left[\frac{4}{9}(A^2-3B)\right].
\end{equation}
Thus we get the total action difference between the black hole and
pure AdS space
\begin{equation}
\Delta I_{bulk}+\Delta I_{GH} =\frac{V(\vec{x})\beta}{16\pi
G_5l^2}\left(-\mu
l^2-\frac{2}{3}(q_1^4+q_2^4+q_3^4-q_1^2q_2^2-q_2^2q_3^2-q_1^2q_3^2)\right).
\end{equation}
The appearance of the non-linear terms of charges in this formula is
due to the asymptotical behavior of the scalar fields. When $\mu=0$,
those terms do not vanish. This is not a reasonable result.  As
argued in Ref.~\cite{counter}, we should add a counterterm $\int
d\tau d^3x{\sqrt{h}\vec{\phi}}^2$  to the action, which just cancels
the part
$-\frac{2}{3}(q_1^4+q_2^4+q_3^4-q_1^2q_2^2-q_2^2q_3^2-q_1^2q_3^2)$.
Finally we arrive at
\begin{equation}
\label{4eq24}
 \Delta I=\frac{V(\vec{x})\beta}{16\pi
G_5l^2}\left(-\mu l^2\right).
\end{equation}
Thus we find that this action difference is always negative, which
means that no Hawking-Page transition happens between the $AdS_5$
black hole and the thermal $AdS_5$ space-time here, and the dual
R-charge field theories are always in the deconfinement phase.

It should be noted here, the counterterm $\int d\tau
d^3x{\sqrt{h}\vec{\phi}}^2$ in the gauged supergravity is just a
special form of counterterms to eliminate the non-linear terms of
charges and divergent terms. There are general counterterms for
general gauged supergravity theories, which have been discussed in
~\cite{Batrachenko:2004fd}. For this 5-dimensional R-charged Ricci
flat AdS black hole, one can find this counterterm $$\int d\tau
d^3x\sqrt {h}(W(\phi)-3/l),$$ where $W(\phi)$ is superpotential, and
we have subtracted the contribution of the gravity counterterm $\int
d\tau d^3x\sqrt{h}3/l$. After substituting the explicit form of
$W(\phi)$ given in~\cite{Batrachenko:2004fd}, one finds the
non-linear charge term is precisely cancelled. This counterterm is
equivalent to the counterterm $\int d\tau d^3x
\sqrt{h}\vec{\phi}^2$. In fact, for some $\phi_0$ we have
$W(\phi_0)=3/l$, so expand $W(\phi)-3/l$ around $\phi_0$, and one
can get expression like $\int d\tau d^3x\sqrt{h}\vec{\phi}^2$. We
will give more detail discussion for this counterterm in the next
section.

Now we turn to the case with an IR cutoff $r_{IR}$.  As in the case
of Schwarzschild-AdS black holes, we introduce $r_{max}=\max[r_0,
r_{IR}]$. The integral of the background starts from $r_{IR}$ to
$r_{uv}$ and the integral of the black hole starts from $r_{max}$ to
$r_{uv}$. We obtain the total action difference after adding the
counterterm
\begin{eqnarray}
\Delta I=\frac{V(\vec{x})\beta}{16\pi G_5l^2}\Bigg{(}\mu
l^2+2r_{IR}^4-2r_{max}^4-\frac{4}{3}Ar_{max}^2-\frac{2}{3}B-\sum_i\frac{2}{3}\frac{\mu
l^2q_i^2}{r_{max}^2+q_i^2}\Bigg{)}.
\end{eqnarray}
When $r_0<r_{IR}$, one has $r_{max}=r_{IR}$, and
\begin{eqnarray}\label{actiondif}
\Delta I=\frac{V(\vec{x})\beta}{16\pi G_5l^2}\Bigg{(}\mu
l^2-\frac{4}{3}Ar_{IR}^2-\frac{2}{3}B-\sum_i\frac{2}{3}\mu
l^2q_i^2\frac{1}{r_{IR}^2+q_i^2}\Bigg{)}\, .
\end{eqnarray}
On the other hand, when $r_0>r_{IR}$, one obtains $r_{max}=r_{0}$,
and
\begin{equation}\label{actiondiff}
\Delta I =\frac{V(\vec{x})\beta}{16\pi G_5l^2}\left(2r_{IR}^4-\mu
l^2\right).
\end{equation}
When $r_{IR}=0$, the action difference reduces to the one
(\ref{4eq24}) without the IR cutoff.

\subsubsection{Phase transition with an IR cutoff}

Here we will discuss the thermodynamics in grand canonical ensemble,
where the chemical potentials and temperature are fixed parameters.
The IR cutoff $r_{IR}$ for this ensemble is a fixed but arbitrary
constant. Since when $r_0$ is large enough $\mu l^2$ becomes very
large and the action difference~(\ref{actiondiff}) becomes a large
negative quantity, the de-confinement phase always exists. Then as
long as the confinement phase exists, a phase transition will
happen. That means to realize a phase transition we only have to
ensure a positive action difference in a certain region of the phase
diagram. Here we give a careful analysis to see whether the IR
cutoff really helps the phase transition, and if it does, what
values should the IR cutoff takes.

Given  some fixed $q_i$'s, there always exists a value of $r_{0}$
which is denoted by $r_{0c}(q_i)$ such that $2r_{0}^4>\mu l^2$ if
$r_0>r_{0c}(q_i)$. This $r_{0c}(q_i)$ always exists because $\mu
l^2$
 approaches
$r_0^4$ when $r_0$ is large enough. Thus we can always find an IR
cutoff $r_{IR}>r_{0c}(q_i)$ which satisfies $2r_{IR}^4-\mu l^2>0$
when $r_0>r_{IR}$. This means that confinement phase always exists
in the $(r_0,q_i)$-space after giving an appropriate IR cutoff, and
this appropriate IR cutoff can always be found.

But we are more interested in whether a confinement phase exists
in the $(T,\Phi_i)$ phase diagram, since the ensemble we are
considering is the grand canonical one.  Note that $\Phi_i$'s
depend on $q_i$'s and $r_0$,
$$(\Phi_i)^2\propto
\frac{q_i^2(r_0^2+q_1^2)(r_0^2+q_2^2)(r_0^2+q_3^2)}{r_0^2(r_0^2+q_i^2)^2}.$$
To keep $\Phi_i$'s unchanged, the charge parameters $q_i$'s have to
change simultaneously when $r_0$ changes. As $r_0\rightarrow\infty$,
$q_i$'s change slowly and tend to fixed values $q_{i}= const. \times
\Phi_{i}$. In other words, fixed chemical potentials are equivalent
to fixed charge parameters when $r_0$ approaches infinity. However,
generally $q_i$'s have an evaluated region, which is denoted by
$\mathcal{Q}$, when $r_0$ changes. This means a fixed chemical
potential $\Phi_i$ corresponds to a set of $q_i$'s. Certainly any
meaningful charge parameter $q_i$ can not be infinity, so
$\mathcal{Q}$ is a bounded region. Now take $r_{0c}$ to be
$\mathrm{max}[r_{0c}(q_i), q_i\in \mathcal{Q}]$. Then from the
discussion in the previous paragraph, we can always find an IR
cutoff $r_{0c}<r_{IR}<r_0$ such that $2r_{IR}^4-\mu l^2>0$. Thus for
any fixed chemical potentials, the confinement phase always exists.
Then we can get to a conclusion that the introduction of an IR
cutoff can realize a positive action difference for a system with
any values of chemical potentials if the value of the IR cutoff is
chosen properly.

Thus we conclude that if the IR cutoff is chosen properly, we can
get a positive action difference for the case $r_0>r_{IR}$
(\ref{actiondiff}) and then to realize a confinement phase. Then we
can say: by introducing the IR cutoff, the action difference
(\ref{actiondiff}) can change its sign, and then the Hawking-Page
transition can occur.  It implies that the confinement-deconfinement
transition can happen for the dual field theory. This means that as
the case without charges, the IR cutoff leads to the existence of
the confinement phase. When temperature is high enough, the
deconfinement transitions happens. In this case, the critical
temperature of transition for the deconfinement transition is
\begin{equation}
T_{c}=\frac{1}{\beta}=\frac{r_0^2r_{IR}^4+2r_0^6-2(q_1^2q_2^2+q_1^2q_3^2+q_2^2q_3^2)r_0^4-4q_1^2q_2^2q_3^2}
{4\pi l^2 r_0^2\sqrt{(r_0^2+q_1^2)(r_0^2+q_2^2)(r_0^2+q_3^2)}}.
\end{equation}
From this critical temperature we find that $r_0$ has a minimum to
assure a positive temperature, but this does not matter the
discussion above.

Since the analytic analysis is not easy to make, in what follows,
we move on to show some phase diagrams in several cases. We should
plot the phase diagrams with chemical potentials and temperatures
as variables and plot out the curve where the phase transition
happens. Besides, we also plot out the phase diagrams in terms of
charge parameters $q$ and horizon radius $r_0$.  These two kinds
of diagrams are equivalent after using the transformation
relations.

\DOUBLEFIGURE[t]{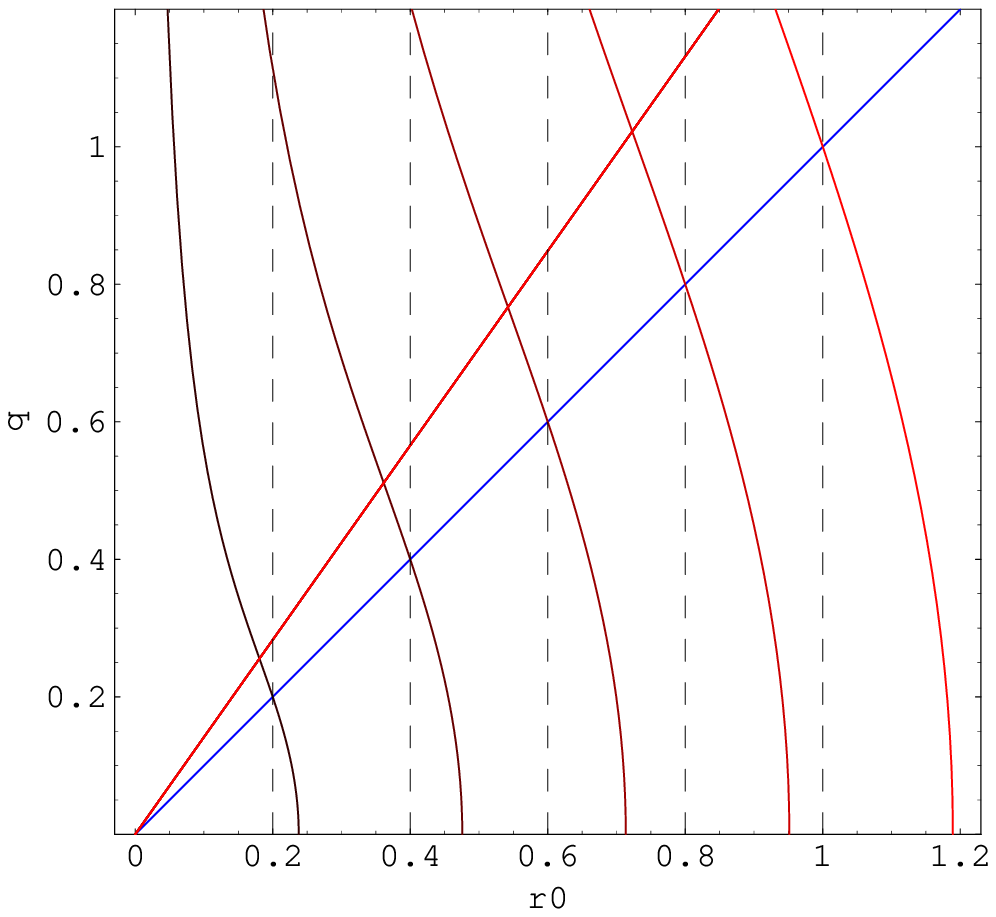, width=.4\textwidth}
{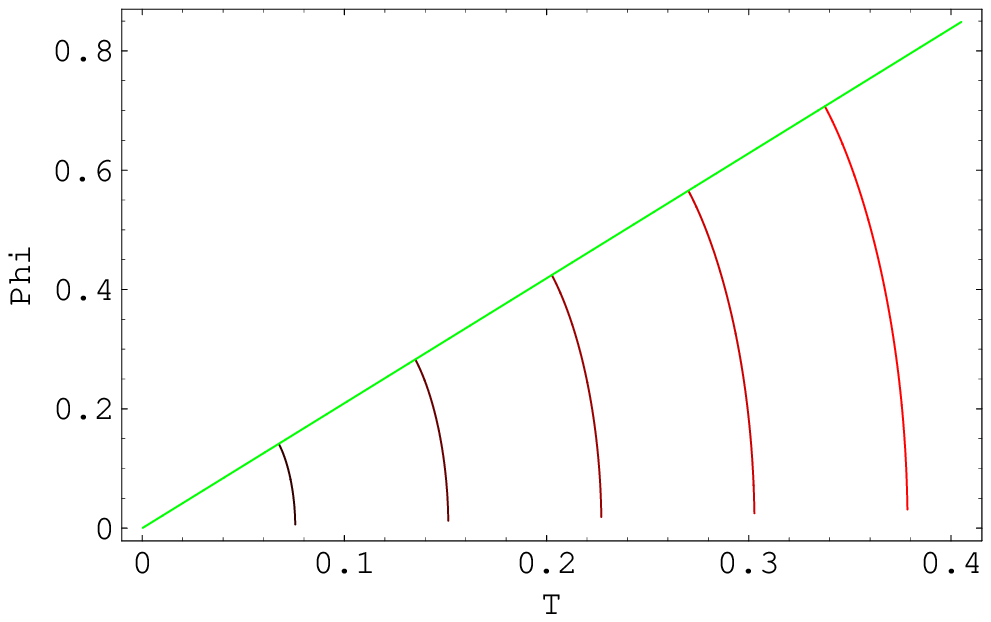, width=.4\textwidth}{$r_0-q$ phase diagram of
5-dimensional R-charged black holes with $q_1=q, q_2=q_3=0$. The
solid curves correspond to the phase transition curves, and each
curve has a fixed IR cutoff $r_{IR}$. With the colors changing from
black to red, the values of $r_{IR}$ increase from $0.2$ to $1.0$
with a step $0.2$. The dashed curves stand for
$r_0=0.2,~0.4,~0.6,~0.8,~1.0$, respectively. The straight blue curve
describes the requirement of $r_0>r_{IR}$, and the straight red
curve divides the diagram into two parts by local stability of
thermodynamics, below which the thermodynamics is local stable.}{
$T-\phi$ phase diagram of 5-dimensional R-charged black hole with
$q_1=q, q_2=q_3=0$. The green curves correspond to the requirement
$r_{IR}<r_0$, here only the stable part is shown. The solid curves
correspond to the phase transition curves, and each curve has a
fixed IR cutoff $r_{IR}$. With the colors changing from black to
red, the values of $r_{IR}$ increase from $0.2$ to $1.0$ with a step
$0.2$.}

\DOUBLEFIGURE[t]{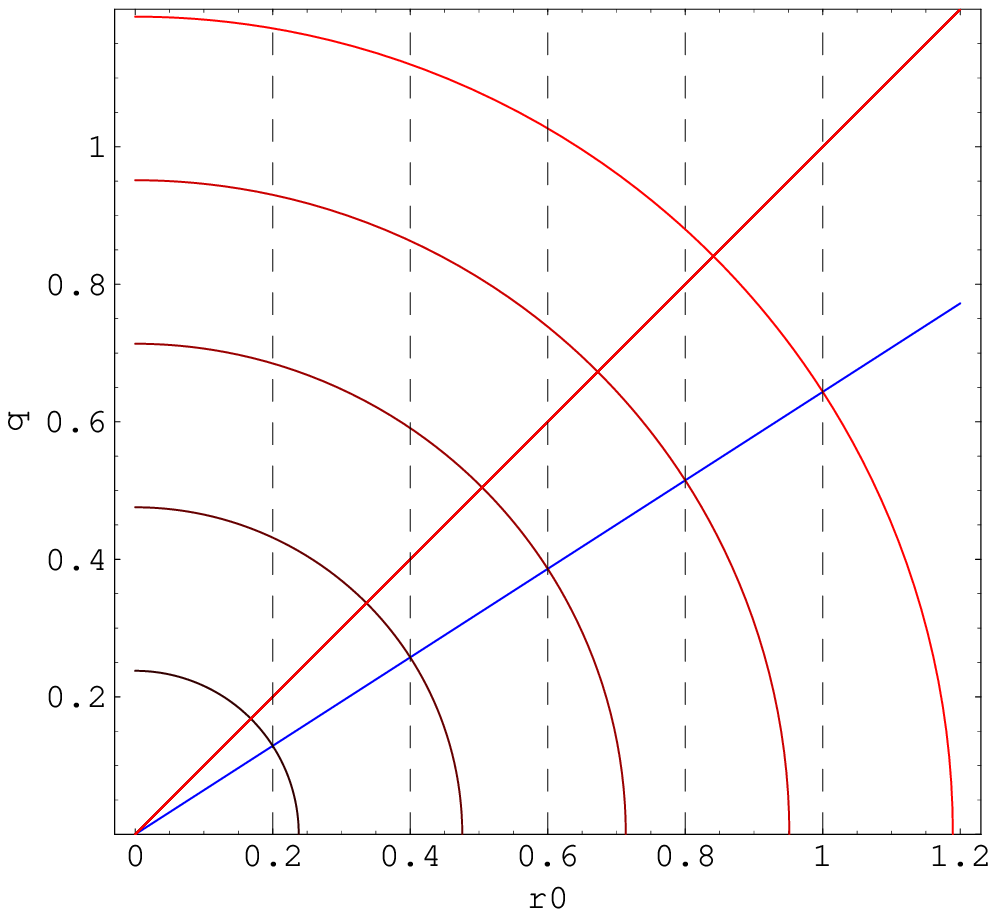, width=.4\textwidth}
{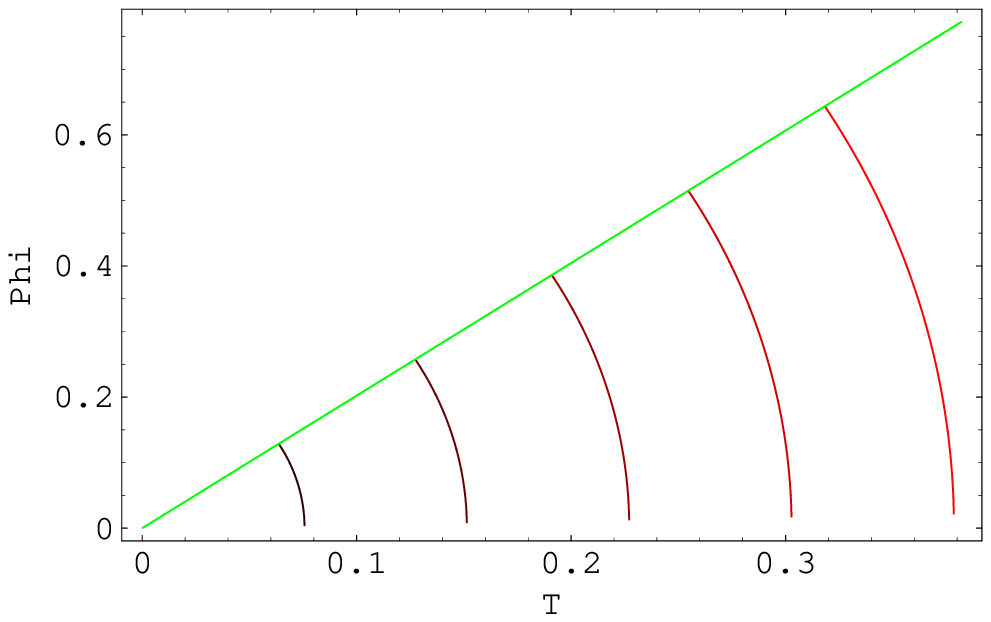, width=.4\textwidth}{$r_0-q$ phase diagram of
5-dimensional R-charged black hole with $q_1= q_2=q, q_3=0$.
}{$T-\phi$ phase diagram of 5-dimensional R-charged black hole
with $q_1= q_2=q, q_3=0$.}

 Figure 1, 3 and 5 are $r_0-q$ phase diagrams of the case
$q_1=q\neq0,~q_2=q_3=0$, $q_1=q_2=q\neq0,~q_3=0$ and
$q_1=q_2=q_3=q\neq0$, respectively. In these figures, the solid
curves correspond to the phase transition curves, across which the
action difference changes its sign, and each curve has a fixed IR
cutoff $r_{IR}$. With the colors changing from black to red, the
values of $r_{IR}$ increase from $0.2$ to $1.0$ with a step $0.2$.
The dashed curves stand for $r_0=0.2,~0.4,~0.6,~0.8,~1.0$. This
means that these straight blue curves describe the requirement of
$r_0>r_{IR}$. Thus, in fact, only the regions below these blue
curves are meaningful for our discussion. From the $r_0-q$
diagrams, one can read out the value of the IR cutoff by the
intersecting points of the blue curves and  the transition curves.

Figure 2, 4 and 6 are $T-\Phi$ diagrams for the case
$\Phi_1=\phi,~\Phi_2=\Phi_3=0$, $\Phi_1=\Phi_2=\phi,~\Phi_3=0$,
and $\Phi_1=\Phi_2=\Phi_3=\phi$, respectively. In these figures,
the green curves correspond to the requirement $r_{IR}<r_0$. With
the color changing from black to red, the values of $r_{IR}$
increase from $0.2$ to $1.0$ with a step $0.2$. In the
paper~\cite{Cai:2007zw}, the charged RN black holes discussed
there are just R-charged AdS black holes with equal R charges.
Here each $T-\phi$ phase diagram is plotted with 5 different
values of $r_{IR}$ to show its influence. The colors of the curves
represent the values of $r_{IR}$, the darker, the smaller.

In the $r_0-q$ diagrams we  also plot the boundary for local
thermodynamic stability~\cite{Cvetic:1999rb, Cvetic:1999ne}. The
straight red curves with $q=\sqrt{2}r_0$, $q=r_0$ and $q=r_0$ in
these $r_0-q$ diagrams correspond to the local thermodynamic
stability curves. The local stability curves are determined by the
Hessian of the Euclidean action
\begin{equation}
I=\beta (E-\Phi_i Q_i)-S,
\end{equation}
with respect to $r_0$ and charge parameters $q_i$'s keeping $\beta$
and $\Phi_i$'s fixed, where $E$ is the mass, $Q_i$'s are physical
charges and $S$ is the entropy of the R-charged black holes. These
thermodynamic quantities can be got from the general thermodynamic
relations
\begin{equation}
E=\left(\frac{\partial I}{\partial
\beta}\right)_{\Phi_i}-\frac{\Phi_i}{\beta}\left(\frac{\partial
I}{\partial\Phi_i }\right)_{\beta}\, ,\quad
S=\beta\left(\frac{\partial I}{\partial \beta}\right)_{\Phi_i}-I\,
,\quad Q_i=-\frac{1}{\beta}\left(\frac{\partial I}{\partial\Phi_i
}\right)_{\beta}\, .
\end{equation}
In this case, the energy will have a constant correction due to the
IR cutoff, while the entropy and physical charges do not change,
\begin{eqnarray}
E&=&\frac{V(\vec{x})}{16\pi G_5 l^2}(3\mu l^2+2r_{IR}^4),\nonumber\\
S&=&\frac{V(\vec{x})}{4
G_5}\sqrt{(r_0^2+q_1^2)(r_0^2+q_2^2)(r_0^2+q_3^2)},\nonumber\\
Q_i&=&\frac{V(\vec{x})}{8\pi
G_5}\frac{q_i}{r_0}\sqrt{(r_0^2+q_1^2)(r_0^2+q_2^2)(r_0^2+q_3^2)}.
\end{eqnarray}
The similar form of these quantities can be found
in~\cite{Cvetic:1999ne, Banerjee:2007by}. The local stability curves
are represented by the red straight curves under which the
thermodynamics is locally stable. In addition, let us mention that
in the $q-r_0$ phase diagrams, only the regions under the blue
curves are physically allowed when the IR cutoff is introduced,
since we are considering the case with $r_0>r_{IR}$. As a result, we
can see from these diagrams that the deconfinement phase transition
always exists and the IR cutoff will not affect the local
thermodynamical stability of the field theories.  In the $T-\phi$
phase diagrams we only plot out the regions, corresponding to the
ones under the blue curves of the $r_0-q$ diagrams.

\DOUBLEFIGURE[t]{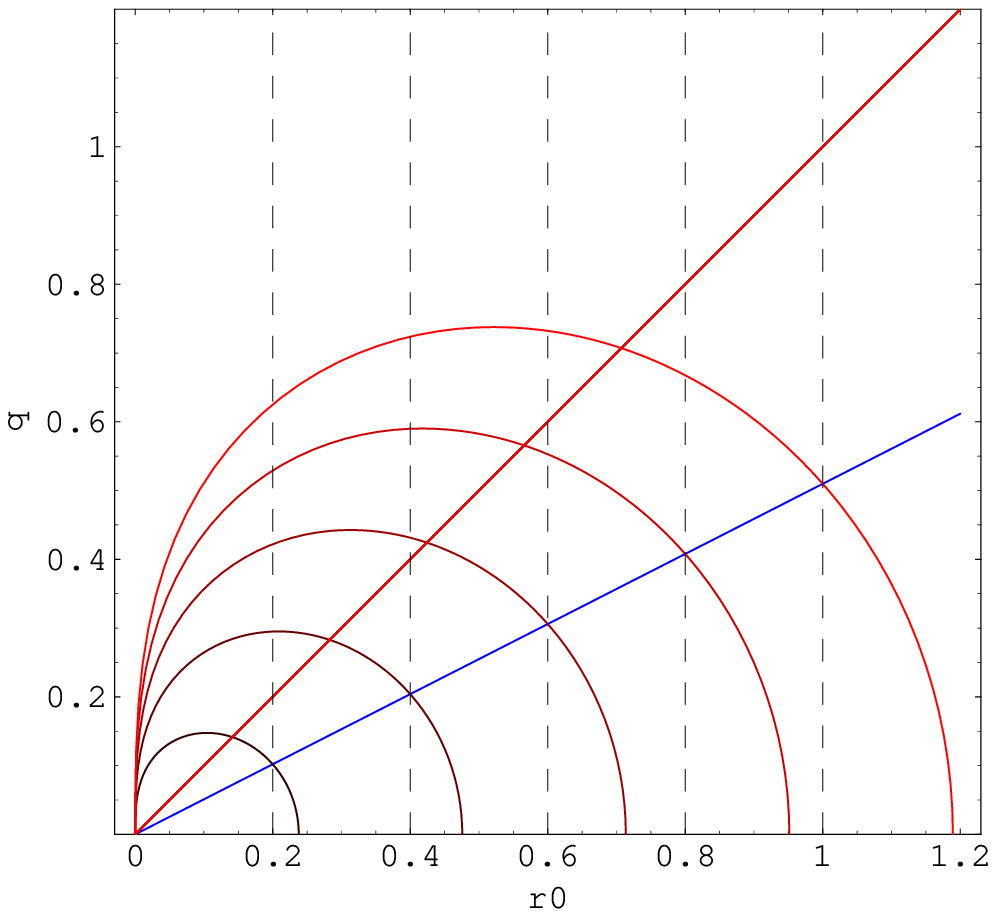, width=.4\textwidth}
{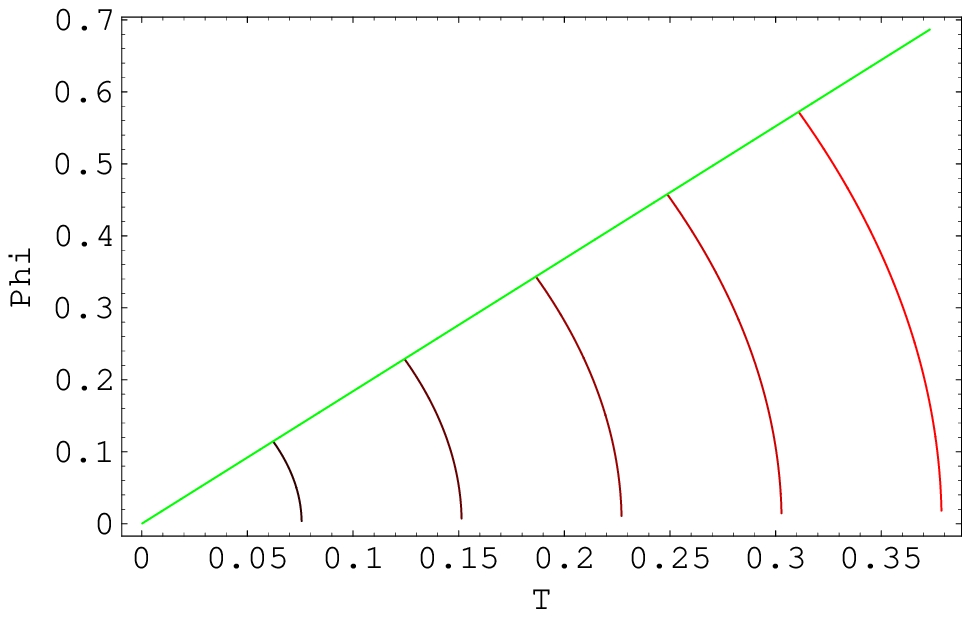, width=.4\textwidth}{$q-r_0$ phase diagram
for the 5-dimensional R-charged black hole with $q_1=
q_2=q_3=q$.}{$\Phi-T$ phase diagram for the 5-dimensional
R-charged black hole with $q_1= q_2=q_3=q$.}

\subsection{R-charged $AdS_4$ black holes}
For rotating M2-branes in 11 dimensional supergravity, there are 8
transverse spatial dimensions. Thus there can be at most 4 angular
momenta. After dimensional reduction, there will be at most 4
charges parameterized by $q_i,i=1,2,3,4$.

The decoupling limit of the rotating black M2-brane after reduction
is four dimensional AdS black hole, which can be written
as~\cite{Cvetic:1999xp}
\begin{equation}
ds^2=-({\mathcal{H}}_1{\mathcal{H}}_2{\mathcal{H}}_3{\mathcal{H}}_4)^{-1/2}fdt^2+
({\mathcal{H}}_1{\mathcal{H}}_2{\mathcal{H}}_3{\mathcal{H}}_4)^{1/2}\left(f^{-1}dr^2+r^2(dx_1^2+dx_2^2)\right)
\end{equation}
where
\begin{equation}
\label{ffuction} f=-\frac{\mu}{r}+\frac{4 r^2}{l^2}
{\mathcal{H}}_1{\mathcal{H}}_2{\mathcal{H}}_3{\mathcal{H}}_4\, ,
\quad {\mathcal{H}}_i =1+\frac{\mu \sinh^2{\beta_i}}{r}\, ,
\end{equation}
\begin{equation}
X_i={\mathcal{H}}_i^{-1}({\mathcal{H}}_1{\mathcal{H}}_2{\mathcal{H}}_3{\mathcal{H}}_4)^{1/4}\,
,\quad A^i_t=\frac{1-{\mathcal{H}}_i^{-1}}{\sinh{\beta_i}}\, .
\end{equation}
Define $ q_i^2=\mu \sinh^2{\beta_i} $, then we have
\begin{equation}
{\mathcal{H}}_i =1+\frac{q_i^2}{r}\, , \quad
A^i_t=\frac{\sqrt{\mu}(1-{\mathcal{H}}_i^{-1})}{q_i}\, .
\end{equation}
The effective action in 4 dimensions is
\begin{equation}
I=-\frac{1}{16\pi G_4}\int \sqrt{-g} \Biggl (R-\frac{1}{2}(\partial
\vec{\varphi})^2+\frac{4}{l^2}\sum_{i<j} X_iX_j
-\frac{1}{4}\sum_{i}X_i^{-2}(F^i)^2\Biggr)\, ,
\end{equation}
The relation between scalars $X_i$ and $\vec{\varphi}=(\varphi_1,
\varphi_2, \varphi_3)$ is given by
\begin{equation}
X_i=e^{-\frac{1}{2} \vec{a}_i\cdot \vec{\varphi}}\, ,
\end{equation}
where the vectors $\vec{a}_i$ are given by
\begin{equation}
\vec{a}_1=(1,1,1),\quad \vec{a}_2=(1,-1,-1),\quad
\vec{a}_3=(-1,1,-1),\quad\vec{a}_4=(-1,-1,1).
\end{equation}

\subsubsection{Euclidian action of $AdS_4$ R-charged black holes}
Here and in the next section we also work in the grand canonical
ensemble where the chemical potentials are fixed at the boundary as
in the case of $AdS_5$ R-charged black holes, and the reference
background is also pure thermal $AdS_4$ spacetime with zero valued
charges but maybe nonzero electric potentials. Then we come to the
calculation of the difference of Euclidean actions of the two
solutions.

Substituting the black hole solution into the action,  we get the
on-shell Euclidean action
\begin{eqnarray}
I&=&\frac{V(\vec{x})}{16\pi G_4}\int d\tau dr\Bigg{[}-\frac{\mu}{2}
\sum_{i=1}^4 \frac{q_i^2}{(r+q_i^2)^2}+\frac{4}{l^2}\left(6r^2+3 A
r+B\right)\Bigg{]}.
\end{eqnarray}
Here we have introduced the following quantities
\begin{equation}
A=\sum_iq_i^2,\quad B=\sum_{i<j}q_i^2q_j^2,\quad
C=\sum_{i<j<k}q_i^2q_j^2q_k^2,\quad D=q_1^2q_2^2q_3^2q_4^2.
\end{equation}
We first consider the case without an IR cutoff. In this case, the
bulk action for the black hole is
\begin{eqnarray}
I^{bl}_{bulk} &=&\frac{V(\vec{x})\beta}{16\pi
G_4}\Bigg{[}\frac{\mu }{2}\sum_{i=1}^4
\left(\frac{q_i^2}{r_{uv}+q_i^2}-\frac{q_i^2}{r_{0}+q_i^2}\right)+
\frac{8}{ l^2}(r^3_{uv}-r^3_0)\nonumber \\
&&+\frac{6}{ l^2} A(r_{uv}^2-r_0^2) + \frac{4}{l^2}
B(r_{uv}-r_0)\Bigg{]}\, ,
\end{eqnarray}
and for the pure $AdS_4$, the bulk action is
\begin{equation}
I^{ba}_{bulk}=\frac{V(\vec{x})\beta'}{16\pi G_4}( 8r^3_{uv}),
\end{equation}
which is also unaffected by the values of electric potentials by the
same reason argued in the case of $AdS_5$ R-charged black holes. The
contributions from the Gibbons-Hawking surface term are
\begin{equation}
I^{bl}_{GH} =-\frac{V(\vec{x})\beta}{8\pi G_4l^2}\left(12
r_{uv}^3+9 A r_{uv}^2+6B
r_{uv}+3C-\frac{3l^2}{2}\mu+\cdots\right),
\end{equation}
and
\begin{equation}
I^{ba}_{GH}=-\frac{1}{8\pi G_4}\int d\tau d^2x
\sqrt{h}K=-\frac{V(\vec{x})\beta'}{8\pi G_4 l^2}\left(12
r_{uv}^3\right),
\end{equation}
respectively, for the black hole solution and pure AdS space, where
from the equation~(\ref{period}), we have the relation between the
two temperatures
\begin{equation}
\beta'=\beta \left.
\sqrt{\frac{({\mathcal{H}}_1{\mathcal{H}}_2{\mathcal{H}}_3{\mathcal{H}}_4)^{\frac{1}{2}}fl^2}{4
r^2}} \right|_{r_{uv}},
\end{equation}
where $\beta$ is the inverse Hawking temperature of the black
hole.  Because there is something subtle in this case, we write
out the relation between $\beta'$ and $\beta$ explicitly as
follows,
\begin{equation}
\beta r_{uv}^3-\beta'r_{uv}^3=\beta \left(-\frac{3}{4}A
r_{uv}^2+\frac{3}{32}(A^2-8B)r_{uv}+\frac{\mu l^2}{8}+
s_1(q_1,q_2,q_3,q_4)\right),
\end{equation}
where there is a non-linear charge term which can be written as
\begin{eqnarray}
s_1(q_1,q_2,q_3,q_4) &=&\frac{1}{128}\left
[-5(q_1^6+q_2^6+q_3^6+q_4^6)+9(q_1^4q_2^2+q_1^4q_2^2+q_1^4q_4^2
\right. \nonumber
\\
&+&q_2^4q_1^2+q_2^4q_3^2+q_2^4q_4^2+q_3^4q_1^2+q_3^4q_2^2+q_3^4q_4^2+q_4^4q_1^2\nonumber\\
&+&q_4^4q_2^2+q_4^4q_3^2)
-54(q_1^2q_2^2q_3^2+q_1^2q_2^2q_4^2+q_1^2q_3^2q_4^2+q_2^2q_3^2q_4^2)].
\end{eqnarray}
Then we find
\begin{eqnarray}
\Delta I_{total}&=&\lim_{r_{uv}\rightarrow
\infty}\frac{V(\vec{x})\beta}{16\pi G_4}\Bigg{[}-\frac{\mu
}{2}\sum_{i=1}^4
\left(\frac{q_i^2}{r_{0}+q_i^2}\right)-\frac{1}{2l^2}
(3A^2-8B)r_{uv}\nonumber
\\&&-16\frac{s_1}{l^2}-6\frac{C}{l^2}+\mu-8\frac{r_0^3}{l^2}-6 \frac{ Ar_0^2}{l^2}-
4\frac{ Br_0}{l^2}+\cdots \Bigg{]}\, .
\end{eqnarray}
Note that $r_0$ is the horizon of the black hole,  satisfying
\begin{equation}
\label{horizonAdS4}
-\frac{\mu}{r_0}+\frac{4r_0^2}{l^2}{\mathcal{H}}_1(r_0){\mathcal{H}}_2(r_0){\mathcal{H}}_3(r_0){\mathcal{H}}_4(r_0)=0,
\end{equation}
we arrive at
\begin{eqnarray}
\label{deltaItotal} \Delta I_{total}&=&\lim_{r_{uv}\rightarrow
\infty}\frac{V(\vec{x})\beta}{16\pi G_4}\Bigg{[}-\mu-\frac{1}{2l^2}
(3A^2-8B)r_{uv}-16\frac{s_1}{l^2}-4\frac{C}{l^2}\Bigg{]}\, .
\end{eqnarray}
One can see that as $r_{uv} \to \infty$, the action  diverges,
unless the four charges are equal or at least equal two by two. This
problem is common in theories with scalar fields. The divergence is
due to the asymptotical behavior of these scalar fields. The similar
problem arises in the so called ``boundary counterterm"
method~\cite{Balasubramanian:1999re, Henningson:1998gx,
Kraus:1999di, deHaro:2000xn, Skenderis:2000in, counter}. In these
references, one can remove the divergence by adding a counterterm
$I^g_{ct}$ into the action,
\begin{equation}
I=I_{bulk}+I_{GH}+I^g_{ct}\, .
\end{equation}
These counterterms are constructed by boundary curvature,
\begin{eqnarray}
I^g_{ct}&=&\frac{1}{8\pi G_d}\int
d^{d-1}x \sqrt{h}\Bigg{[}(d-2)/l+\frac{l}{2(d-3)}\mathcal{R}\nonumber\\
&&+\frac{l^3}{2(d-3)^2(d-5)}\left({\mathcal{R}}^{ab}{\mathcal{R}}_{ab}-\frac{d-1}{4(d-2)}{\mathcal{R}}^2\right)+\cdots\Bigg{]}\,
,
\end{eqnarray}
where $\mathcal{R}, {\mathcal{R}}_{ab}$ are Ricci scalar and Ricci
tensor of the boundary. Thus one can call them gravity counterterms,
and denote the sum by an index $g$. However, for theories with
scalar fields, the divergence can not be eliminated even after one
has added the gravity counterterm. To eliminate the divergence,
generalized counterterms for the theories with scalars should be
added as follows.
\begin{eqnarray}
I_{ct}&=&\frac{1}{8\pi G_d}\int
d^{d-1}x \sqrt{h}\Bigg{[}W(\phi)+\frac{l}{2(d-3)}\mathcal{R}\nonumber\\
&&+\frac{l^3}{2(d-3)^2(d-5)}\left({\mathcal{R}}^{ab}{\mathcal{R}}_{ab}-\frac{d-1}{4(d-2)}{\mathcal{R}}^2\right)+\cdots\Bigg{]}\,
,
\end{eqnarray}
where $W(\phi)$ is the superpotential and $\mathcal{R},
{\mathcal{R}}_{ab}$ are the Ricci scalar and Ricci tensor of the
boundary.

This kind of counterterm was first derived
in~\cite{Skenderis:0105276} for the domain wall solution in five
dimensional supergravity, and the
subsequent~\cite{Skenderis:0112119} for a more complete derivation.
Here, the superpotential counterterm is by no means the only one
needed. In general one also needs counterterms involving derivatives
of scalars. By using Hamiltonian/Hamilton-Jacobi methods, the
general analysis for gravity coupled to scalars with the complete
set of counterterms has been given~\cite{Skenderis:0404176}. And
more information about the counterterm of the system with scalar
fields coupling to gravity can be found in~\cite{deHaro:2000xn,
Skenderis:0404176,Skenderis:0505190,Batrachenko:2004fd}\footnote{We
would like thank Kostas Skenderis for useful comments on this
point.}.

Here, since we are interested in the cases of Ricci flat black
holes, this boundary counterterm is fully determined by the
superpotential
\begin{eqnarray}
I_{ct}&=&\frac{1}{8\pi G_d}\int d^{d-1}x \sqrt{h} W(\phi),
\end{eqnarray}
for any dimension. Certainly, with this counterterm, one can give
appropriate Euclidean action for the black holes without considering
the procedure of selecting a proper background. However, here since
we are discussing the possible Hawking-Page phase transitions
between the black hole and the background spacetime, it is more
natural to use the background subtraction method. Thus in what
follows we will subtract the contribution of the pure gravity
counterterm, which means that the counterterm should be
\begin{equation}
\label{gcounterterm} I^s_{ct}=\frac{1}{8\pi G_d}\int d^{d-1}x
\sqrt{h}\left(W(\phi)-(d-2)/l\right).
\end{equation}
For the $D=4$ R-charged black hole, we have (noting the AdS scale
$l$ in the function $f$ of~(\ref{ffuction}) is different from the
standard one by a factor ``1/2", so in the following calculation we
have to change $l$ in~(\ref{gcounterterm}) to be $l/2$)
\begin{equation}
W(\phi)=\frac{1}{l}\sum_{i}X_i\, ,\quad X_i=e^{-\frac{1}{2}\vec{a}_i
\cdot \vec{\phi}}
\end{equation}
Thus, the counterterm for this four dimensional R-charged black hole
becomes
\begin{equation}
I^s_{ct}=\frac{1}{8\pi G_4l}\int d\tau d^{2}x
\sqrt{h}\left(\sum_{i}X_i-4\right).
\end{equation}
It is easy to find that the integrand in the above equation has the
following expansion
\begin{equation}
\label{counterexpand}
\frac{1}{l}\sqrt{h}\left(\sum_{i}X_i-4\right)=\frac{1}{4l^2}
(3A^2-8B)r_{uv}-\frac{1}{16l^2} s_2(q_1,q_2,q_3,q_4)+\cdots,
\end{equation}
where there are non-linear charge terms like the one
in~(\ref{deltaItotal}), which is denoted by
\begin{eqnarray}
s_2(q_1,q_2,q_3,q_4)&=&5(q_1^6+q_2^6+q_3^6+q_4^6)-9(q_1^4q_2^2+q_1^4q_2^2+q_1^4q_4^2\nonumber
\\
&+&q_2^4q_1^2+q_2^4q_3^2+q_2^4q_4^2+q_3^4q_1^2+q_3^4q_2^2+q_3^4q_4^2+q_4^4q_1^2\nonumber\\
&+&q_4^4q_2^2+q_4^4q_3^2)
+22(q_1^2q_2^2q_3^2+q_1^2q_2^2q_4^2+q_1^2q_3^2q_4^2+q_2^2q_3^2q_4^2)
.
\end{eqnarray}
Note that the first term in~(\ref{counterexpand}) precisely cancels
the divergence term in the action difference~(\ref{deltaItotal}),
while the second term in~(\ref{counterexpand}) exactly remove the
non-linear charge terms by following relation
\begin{equation}
16 s_1+4 C=-\frac{1}{8} s_2\, ,
\end{equation}
so after considering this counterterm, we finally get the Euclidean
action difference between the black hole and pure AdS background
\begin{equation}
\Delta I=-\frac{V(\vec{x})\beta}{16\pi G_4}\mu<0,
\end{equation}
which means that there is no phase transition in this case, and the
black hole solution dominates and the dual field theory is in the
deconfinement phase.

Now  we turn to the case with an IR cutoff $r_{IR}$. In this case
the contributions from the Gibbons-Hawking surface term and the
counterterm which are calculated on the UV boundary are not
affected, and the bulk part changes to
\begin{eqnarray}
I^{bl}_{bulk}&=&\frac{V(\vec{x})\beta}{16\pi G_4}\Bigg{[}\frac{\mu
}{2}\sum_{i=1}^4
\left(\frac{q_i^2}{r_{uv}+q_i^2}-\frac{q_i^2}{r_{max}+q_i^2}\right)+
\frac{8}{ l^2}(r^3_{uv}-r^3_{max})\nonumber \\
&&+\frac{6}{ l^2} A(r_{uv}^2-r_{max}^2) + \frac{4}{l^2}
B(r_{uv}-r_{max})\Bigg{]}\, ,
\end{eqnarray}
where we have introduced $r_{max}=\mathrm{max}[r_0, r_{IR}]$. The
action of the background becomes
\begin{equation}
I^{ba}_{bulk}=\frac{V(\vec{x})\beta'}{16\pi
G_4l^2}(8r_{uv}^3-8r_{IR}^3).
\end{equation}
Considering the contributions from
 the Gibbons-Hawking  surface terms and counterterms, we obtain the total action
difference
\begin{eqnarray}
\Delta I&=&\frac{V(\vec{x})\beta}{16\pi
G_4}\Big{(}-\mu+\frac{8}{l^2}r_{IR}^3+\frac{\mu }{2}\sum_{i=1}^4
\left(\frac{q_i^2}{r_{0}+q_i^2}\right)-\frac{\mu }{2}\sum_{i=1}^4
\left(\frac{q_i^2}{r_{max}+q_i^2}\right)\nonumber\\
&&+\frac{8}{l^2}r_0^3-\frac{8}{l^2}r_{max}^3+\frac{6}{l^2}Ar_{0}^2-\frac{6}{l^2}Ar_{max}^2+\frac{4}{l^2}Br_0-\frac{4}{l^2}Br_{max}\Big{)}.
\end{eqnarray}
When $r_0<r_{IR}$, one should have $r_{max}=r_{IR}$, and
\begin{eqnarray}
\Delta I&=&\frac{V(\vec{x})\beta}{16\pi
G_4}\Big{(}-\mu+\frac{8}{l^2}r_{IR}^3+\frac{\mu }{2}\sum_{i=1}^4
\left(\frac{q_i^2}{r_{0}+q_i^2}\right)-\frac{\mu }{2}\sum_{i=1}^4
\left(\frac{q_i^2}{r_{IR}+q_i^2}\right)\nonumber\\
&&+\frac{8}{l^2}r_0^3-\frac{8}{l^2}r_{IR}^3+\frac{6}{l^2}Ar_{0}^2-\frac{6}{l^2}Ar_{IR}^2+\frac{4}{l^2}Br_0-\frac{4}{l^2}Br_{IR}\Big{)}.
\end{eqnarray}
On the other hand, when $r_0>r_{IR}$, we obtain $r_{max}=r_{0}$.
Considering (\ref{horizonAdS4}), we have a simple expression for
the action difference
\begin{equation}
\Delta I=\frac{V(\vec{x})\beta}{16\pi
G_4}\left(-\mu+\frac{8}{l^2}r_{IR}^3\right).
\end{equation}
This is just the one we want. When $r_{IR} \to 0$, the action
reduces to the case without an IR cutoff.

\subsubsection{Phase transition with an IR cutoff}

Next we analyze the phase structure for the case with $r_0
>r_{IR}$. From equation (\ref{horizonAdS4}) we find that
$$\mu=\frac{4}{l^2}r_0^3
{\mathcal{H}}_{1}(r_0){\mathcal{H}}_{2}(r_0){\mathcal{H}}_{3}(r_0){\mathcal{H}}_{4}(r_0)$$
approaches to $\frac{4}{l^2}r_0^3$ when $r_0$ is large enough.
Thus as argued in the five dimensional case, for any values of
$q_i$'s, there always exists a value of $r_{0}$ which is denoted
by $r_{0c}(q_i)$ such that $\frac{8}{l^2}r_{0}^3>\mu$ if
$r_0>r_{0c}(q_i)$. This $r_{0c}(q_i)$ always exists due to the
properties of $\mu$. Therefore, we can always find an IR cutoff
$r_{0c}(q_i)<r_{IR}<r_0$ which satisfies
$\frac{8}{l^2}r_{IR}^3-\mu>0$. The latter indicates a confinement
phase. This means the confinement phase always exists in the
$(r_0,q_i)$-space once an appropriate IR cutoff is given.

On the other hand, when $\mu >\frac{8}{l^2}r_{IR}^3$, the action
difference turns to be negative. In this case, the black hole
solution dominates and the dual field theory is in the
deconfinement phase. Therefore when $\mu$ crosses
$\frac{8}{l^2}r_{IR}^3$, the Hawking-Page (deconfinement) phase
transition happens.

Figure 7, 9, 11 and 13 plot the $r_0-q$ phase diagrams for the case
$q_1=q,q_2=q_3=q_4=0$, $q_1=q_2=q,q_3=q_4=0$, $q_1=q_2=q_3=q,
q_4=0$, and $q_1=q_2=q_3=q_4=q$, respectively. The five solid curves
correspond to the phase transition curves, and each curve has a
fixed IR cutoff $r_{IR}$. With the colors changing from black to
red, the values of $r_{IR}$ increase from $0.2$ to $1.0$ with a step
$0.2$. The dashed curves stand for $r_0=0.2,~0.4,~0.6,~0.8$, and
$1.0$, respectively.  The blue curves represent the requirement of
$r_0>r_{IR}$. In these figures, the red curves which start from the
origin correspond to $2q^2=3r_0$, $q^2=r_0$, $q^2=r_0$ and
$q^2=r_0$, respectively. The thermodynamics is local stable in the
region under those red curves.  These curves are determined by the
Hessian of the Euclidean action with respect to $r_0$ and $q_i$ with
$\beta$ and $\Phi_i$ fixed.  Since the regions below these blue
curves satisfy the requirement with $r_0>r_{IR}$, therefore those
regions are always local thermodynamical stable.

Figure 8, 10, 12 and 14 plot the $T-\phi$ phase diagrams for the
 case of $\Phi_1= \phi,~\Phi_2=\Phi_3=\Phi_4=0$,
$\Phi_1=\Phi_2=\phi,~\Phi_3=\Phi_4=0$,
$\Phi_1=\Phi_2=\Phi_3=\phi,~\Phi_4=0$, and
$\Phi_1=\Phi_2=\Phi_3=\Phi_4=\phi$, respectively. The green curves
correspond to the requirement $r_{IR}<r_0$. With the color
changing from black to blue, the values of $r_{IR}$ increase from
$0.2$ to $1.0$ with a step $0.2$. Again, we only plot the region
satisfying the requirement $r_{IR}<r_0$.

\DOUBLEFIGURE[t]{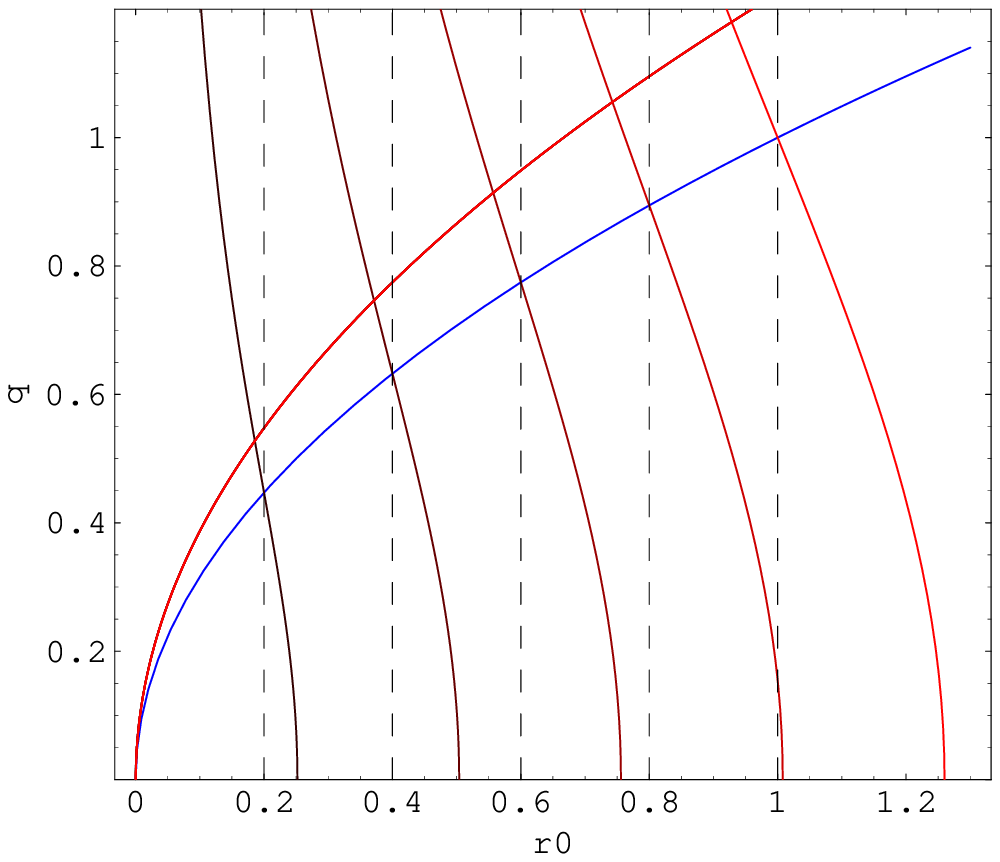, width=.4\textwidth}
{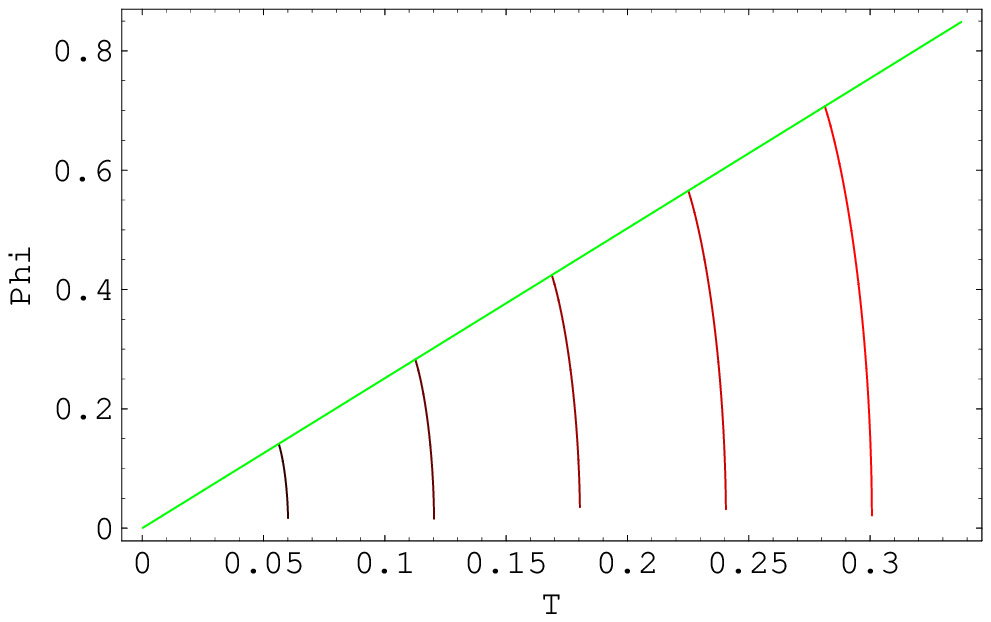, width=.4\textwidth}{$r_0-q$ phase diagram of
4-dimensional R-charged black hole with $q_1=q,
q_2=q_3=q_4=0$.}{$T-\phi$ phase diagram of 4-dimensional R-charged
black hole with $q_1=q, q_2=q_3=q_4=0$.}

\DOUBLEFIGURE[t]{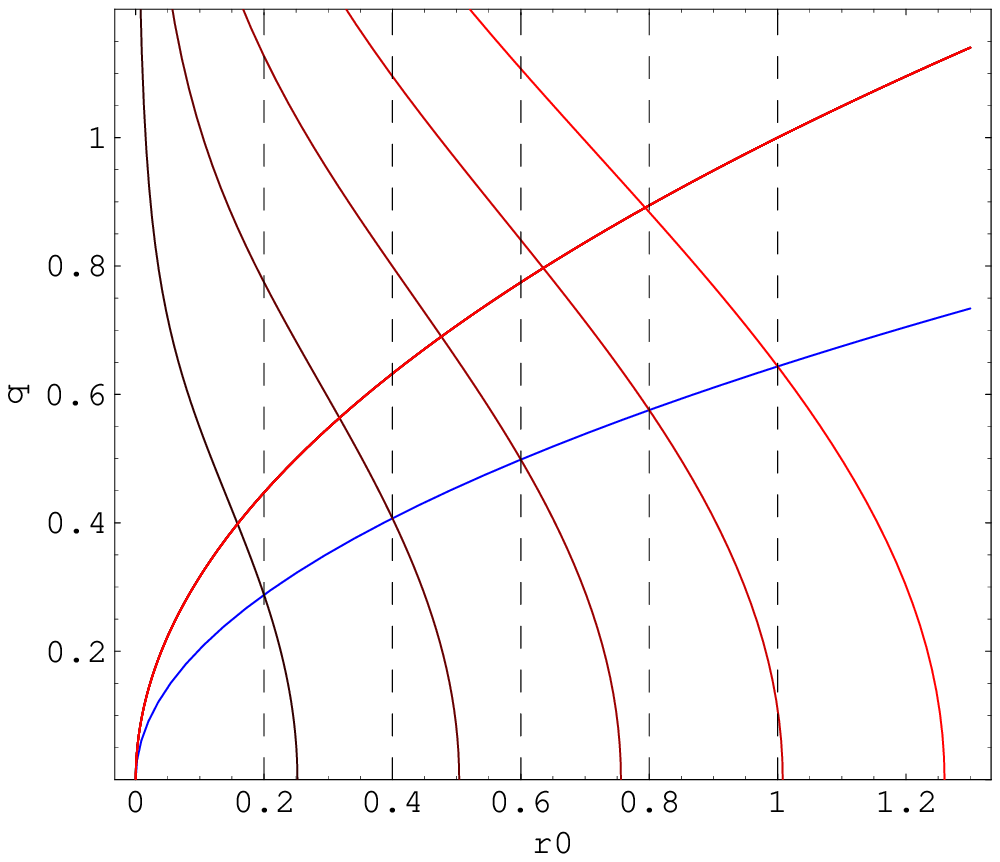, width=.4\textwidth}
{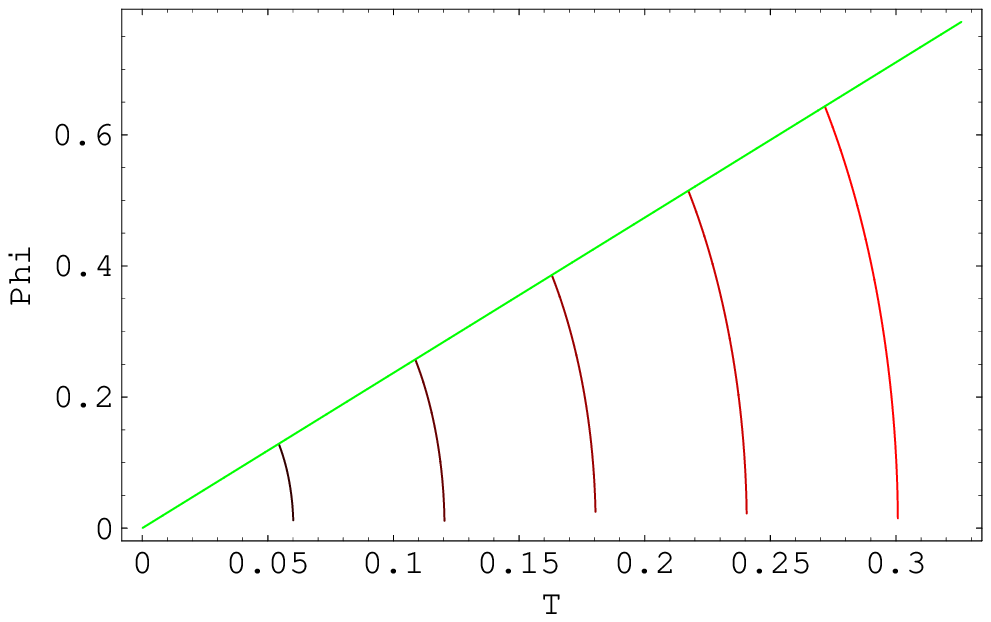, width=.4\textwidth}{$r_0-q$  phase diagram
of 4-dimensional R-charged black hole with $q_1= q_2=q,
q_3=q_4=0$.}{$T-\phi$ phase diagram of 4-dimensional R-charged
black hole with $q_1= q_2=q, q_3=q_4=0$.}

\DOUBLEFIGURE[t]{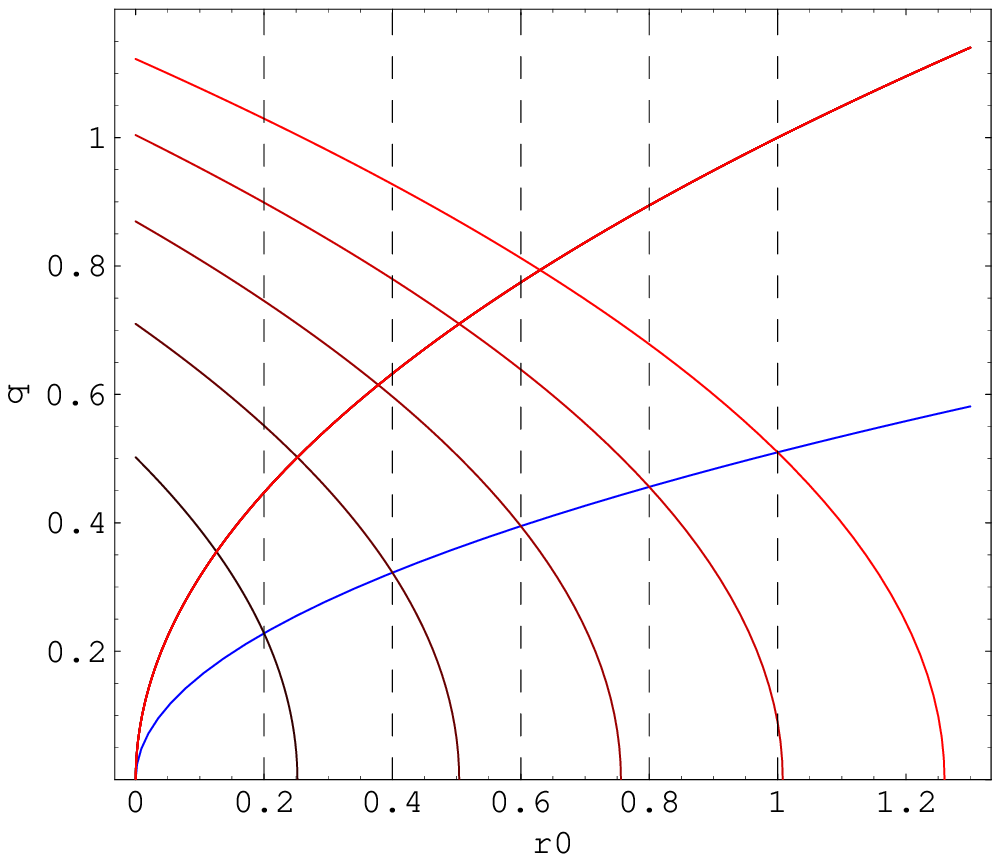, width=.4\textwidth}
{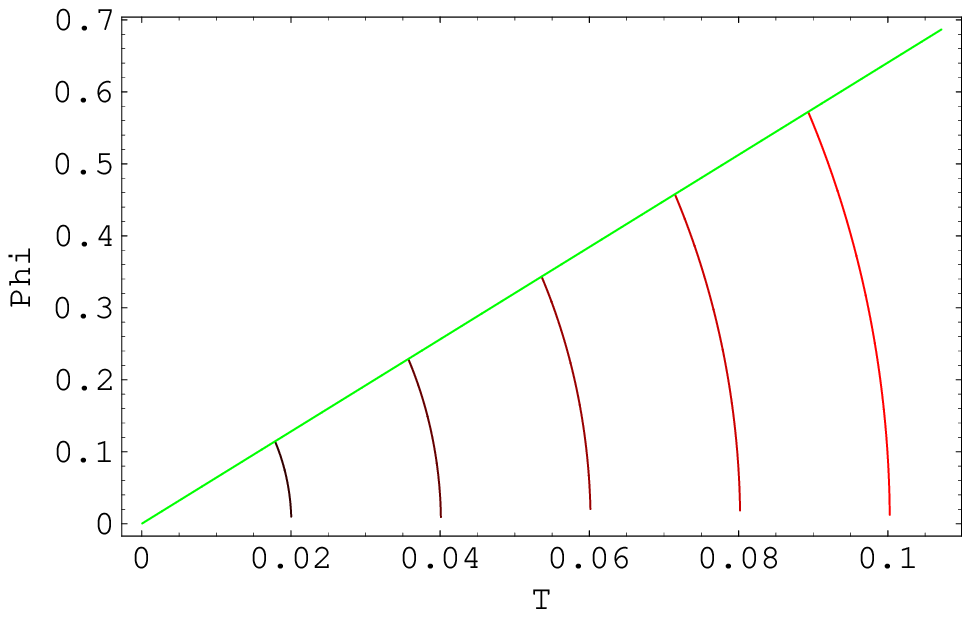, width=.4\textwidth}{$r_0-q$  phase diagram
of 4-dimensional R-charged black hole with $q_1=
q_2=q_3=q,q_4=0$.}{$T-\phi$ phase diagram of 4-dimensional
R-charged black hole with $q_1= q_2=q_3=q,q_4=0$.}

\DOUBLEFIGURE[t]{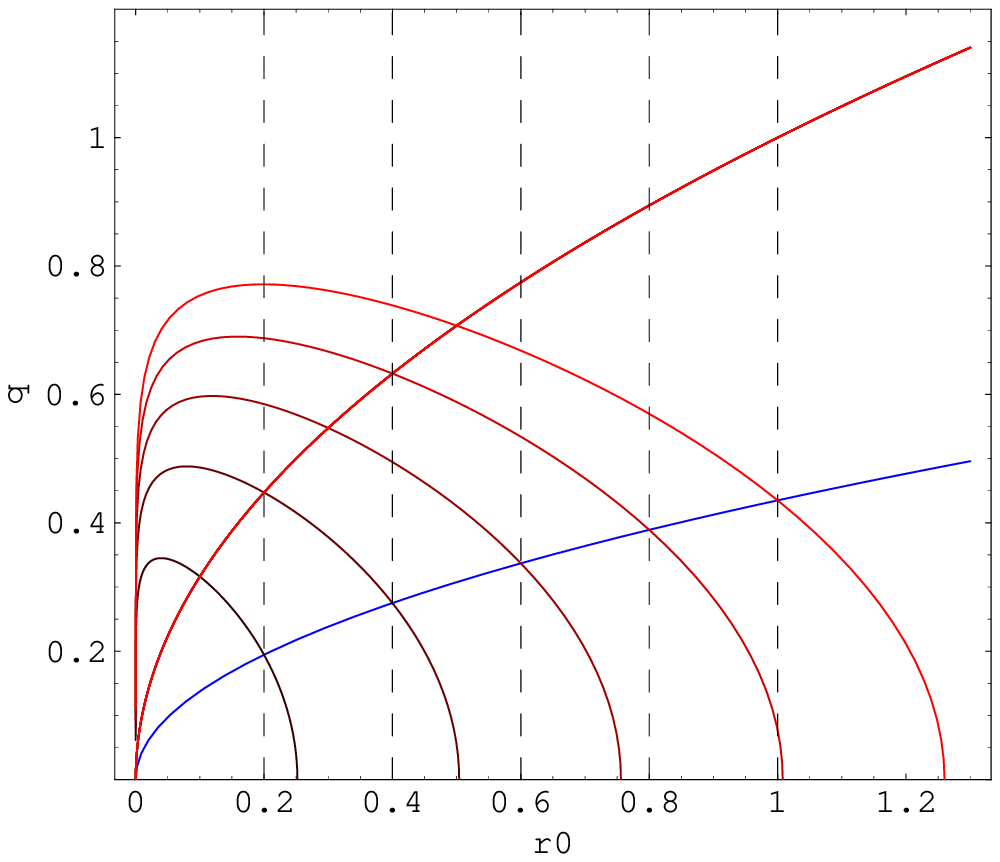, width=.4\textwidth}
{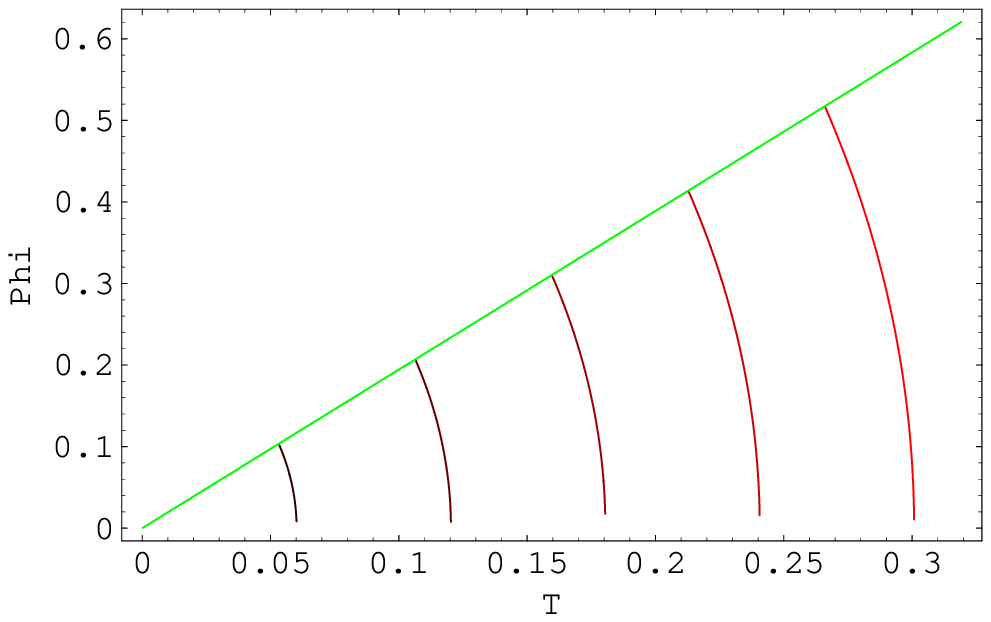, width=.4\textwidth}{ $r_0-q$  phase diagram
of 4-dimensional R-charged black hole with $q_1=
q_2=q_3=q_4=q$.}{$T-\phi$ phase diagram of 4-dimensional R-charged
black hole with $q_1= q_2=q_3=q_4=q$.}

\subsection{R-charged $AdS_7$ black holes}
The R-charged $AdS_7$ black holes have at most two charges
parameterized by $q_1$ and $q_2$. The solution can be written out by
dimensional reduction from 11 dimensional rotating black M5 branes
under decoupling limit~\cite{Cvetic:1999xp}
\begin{eqnarray}
&&ds_7^2=-({\mathcal{H}}_1{\mathcal{H}}_2)^{-\frac{4}{5}}fdt^2
+({\mathcal{H}}_1{\mathcal{H}}_2)^{\frac{1}{5}}\left(f^{-1}dr^2
+r^2d\vec{x}^2\right), \nonumber \\
&&X_i={\mathcal{H}}_i^{-1}({\mathcal{H}}_1{\mathcal{H}}_2)^{\frac{2}{5}},\nonumber
\\
&&f=\frac{r^2}{4l^2}{\mathcal{H}}_1{\mathcal{H}}_2-\frac{\mu}{r^4},\nonumber
\\
&&A^i_t=\frac{\sqrt{\mu}(1-{\mathcal{H}}_i^{-1})}{4lq_i}.
\end{eqnarray}
The effective action in 7 dimensions is
\begin{equation}
I=-\frac{1}{16\pi G_7}\int
d^7x\sqrt{-g}\left(R-\frac{1}{2}(\partial\vec{\varphi})^2
-\frac{V}{l^2}-\frac{1}{4}\sum_{i=1}^2X_i^{-2}(F^i)^2\right),
\end{equation}
where
\begin{equation}
V=-4X_1X_2 -2X_1^{-1}X_2^{-2}- 2X_2^{-1}X_1^{-2} + \frac{1}{2}
(X_1X_2)^{-4},
\end{equation}
\begin{equation}
X_i=e^{-\frac{1}{2}\vec{a}_i\cdot\vec{\varphi}},
\end{equation}
with
\begin{equation}
\vec{a}_0=\left(0,-4 \sqrt{\frac{2}{5}}\right),
\vec{a}_1=\left(\sqrt{2},\sqrt{\frac{2}{5}}\right),
\vec{a}_2=\left(-\sqrt{2},\sqrt{\frac{2}{5}}\right).
\end{equation}
For convenience we also define $A=q_1^2+q_2^2$ and $B=q_1^2q_2^2.$

\subsubsection{Euclidean action of $AdS_7$ R-charged black holes}
Now we come to the calculation of the difference of Euclidean
actions of the R-charged black holes and the pure thermal $AdS_7$
background spacetime. After Euclidean continuation, this solution
becomes
\begin{eqnarray}
&& ds^2 =({\mathcal{H}}_1{\mathcal{H}}_2)^{-\frac{4}{5}}fd\tau^2
+({\mathcal{H}}_1{\mathcal{H}}_2)^{\frac{1}{5}}\left(f^{-1}dr^2
+r^2d\vec{x}^2\right), \\
&& A^i=-i\frac{\sqrt{\mu}(1-{\mathcal{H}}_i^{-1})}{4lq_i}.
\end{eqnarray}
And the Euclidean action
\begin{equation}
I_{Euc}=-\frac{1}{16\pi G_7}\int
d^7x\sqrt{g}\left(R-\frac{1}{2}(\partial\vec{\varphi})^2
-\frac{V}{l^2}-\frac{1}{4}\sum_{i=1}^2X_i^{-2}(F^i)^2\right).
\end{equation}
To avoid the conical singularity in the Euclidean sector of the
black hole solution, the coordinate $\tau$ should get a period
\begin{equation}
\beta=\frac{4\pi}{\left(({\mathcal{H}}_1{\mathcal{H}}_2)^{-\frac{1}{2}}f'(r)
\right)|_{r=r_0}},
\end{equation}
where $r_0$ corresponds to the horizon, and is the largest real root
of $f(r)=0$, i.e.,
\begin{equation}
\label{horizonAdS7}
\mu=\frac{r_0^6}{4l^2}{\mathcal{H}}_{1}(r_0){\mathcal{H}}_{2}(r_0).
\end{equation}
$\beta$ is the inverse Hawking temperature of the black hole.  The
on-shell actions for the black hole and the pure AdS background
\begin{eqnarray}
I^{bl}_{bulk}&=&\frac{V(\vec{x})4^5l^3\beta}{16\pi G_7}
\Bigg{[}\frac{4}{5}l^4q_1^2q_2^2\left(\frac{1}{r_{uv}^2}-\frac{1}{r_{0}^2}\right)
+\frac{3}{20}l^2(q_1^2+q_2^2)(r_{uv}^2-r_0^2)\nonumber
\\&&+\frac{1}{64}(r_{uv}^6-r_0^6)
+\frac{2l^4\mu}{5}\sum_{i=1,2}\left(\frac{q_i^2}{16l^2q_i^2+r_{uv}^4}
-\frac{q_i^2}{16l^2q_i^2+r_0^4}\right)\Bigg{]},
\end{eqnarray}
and
\begin{equation}
I^{ba}_{bulk}=\frac{V(\vec{x})4^5l^3\beta'}{16\pi G_7}
\left(\frac{1}{64}r_{uv}^6\right),
\end{equation}
respectively. From equation~(\ref{period}), the Euclidean time
period of the AdS space is fixed by
\begin{equation}
\beta'=\left.
\beta\sqrt{\frac{4l^2({\mathcal{H}}_1{\mathcal{H}}_2)^{\frac{1}{5}}f}{r^2}}\right|_{r_{uv}}.
\end{equation}
Furthermore, by explicit calculations one can show that the
Gibbons-Hawking surface term and the counterterm discussed in
previous section both have no contribution to this action
difference. As a result the total action difference is
\begin{equation}
\label{action3} \Delta I=-\frac{V(\vec{x})4^5l^6\beta}{16\pi G_7}
\left(\frac{\mu}{32l}\right).
\end{equation}
This is always negative, so there is no Hawking-Page phase
transition in this case.

When an IR cutoff is introduced, the on-shell action of the black
hole becomes
\begin{eqnarray}
I_{bl}&=&\frac{V(\vec{x})4^5l^3\beta}{16\pi G_7 } \left
[\frac{4}{5}l^4q_1^2q_2^2\left(\frac{1}{r_{uv}^2}-\frac{1}{r_{max}^2}\right)
+\frac{3}{20}l^2(q_1^2+q_2^2)(r_{uv}^2-r_{max}^2) \right.
\nonumber
\\&&+ \left. \frac{1}{64}(r_{uv}^6-r_{max}^6)
+\frac{32}{5}l^5U_H^3\sum_{i=1,2}\left(\frac{q_i^2}{16l^2q_i^2
+r_{uv}^4}-\frac{q_i^2}{16l^2q_i^2+r_{max}^4}\right)\right ],
\end{eqnarray}
where $r_{max}=\mathrm{max}[r_0,r_{IR}]$, while for the AdS
background, one has
\begin{equation}
I_{ba}=\frac{V(\vec{x})4^5l^3\beta'}{16\pi G_7 }
\left(\frac{1}{64}(r_{uv}^6-r_{IR}^6)\right).
\end{equation}
Thus the action difference is
\begin{eqnarray}
\Delta I&=&\frac{V(\vec{x})4^5l^3\beta}{16\pi G_7} \Bigg{[}\frac{\mu
l^2}{32}+\frac{1}{64}r_{IR}^6-\frac{1}{64}r_{max}^6-\frac{3}{20}l^2Ar_{max}^2\nonumber
\\&&-\frac{4Bl^4}{5r_{max}^2}
-\frac{2l^4\mu}{5}\sum_{i=1}^2\frac{q_i^2}{r_{max}^4+q_i^216l^2}\Bigg{]}.
\end{eqnarray}
If $r_{0}<r_{IR}$, one has $r_{max}=r_{IR},$ and
\begin{eqnarray}
\Delta I&=&\frac{V(\vec{x})4^5l^3\beta}{16\pi G_7}
\Bigg{[}\frac{\mu
l^2}{32}-\frac{3l^2A}{20}r_{IR}^2-\frac{4Bl^4}{5r_{IR}^2}
-\frac{2l^4\mu}{5}\sum_{i=1}^2\frac{q_i^2}{r_{IR}^4+q_i^216l^2}\Bigg{]}.
\end{eqnarray}
When $r_{0}>r_{IR}$, we should have $r_{max}=r_{0}$, and the
action difference becomes
\begin{equation}
\label{4eq81}
 \Delta I=\frac{V(\vec{x})4^5l^3\beta}{16\pi G_7}
\left ( \frac{1}{64}r_{IR}^6-\frac{\mu l^2}{32}\right ).
\end{equation}
This action difference will reduce to the case without IR cutoff
(\ref{action3}) if the cutoff parameter $r_{IR}$ vanishes. It should
be noted here, for this R-charged black hole, it is easy to find the
counterterm~(\ref{gcounterterm}) does not give any contribution to
the Euclidean action. This is different from the cases in 4 and 5
dimensions.

\subsubsection{Phase transition with an IR cutoff}
\DOUBLEFIGURE[t]{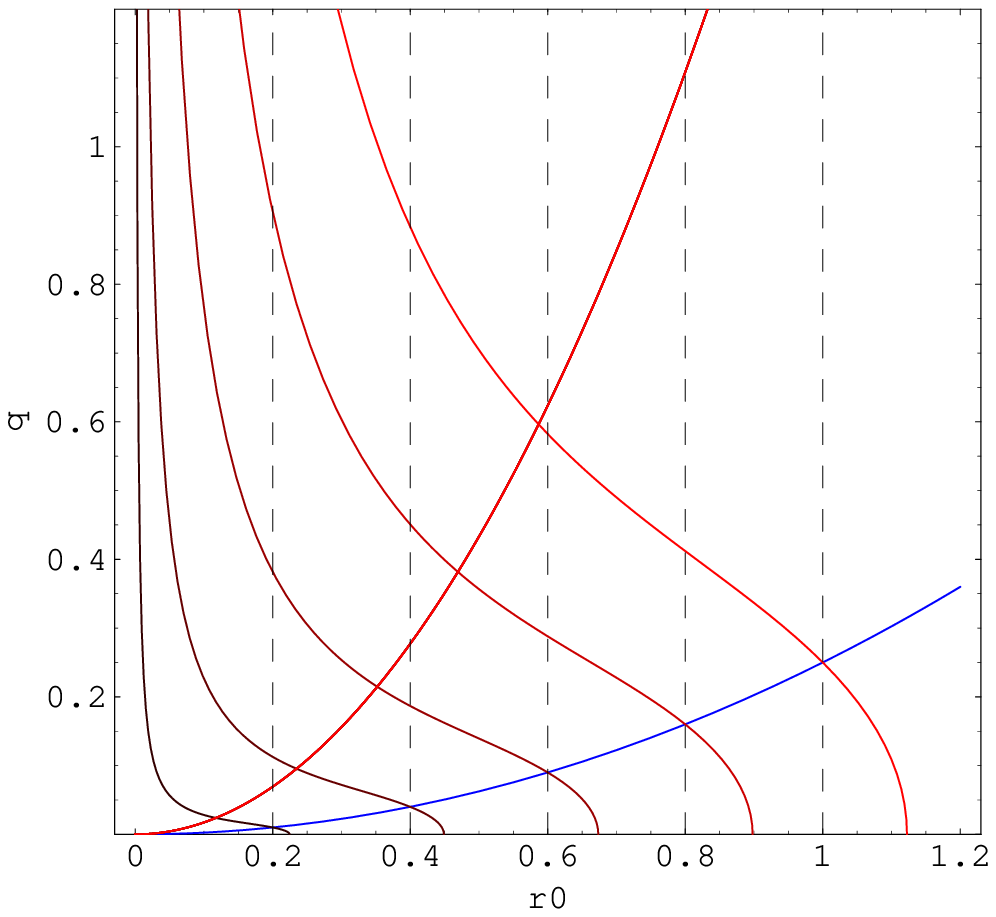, width=.4\textwidth}
{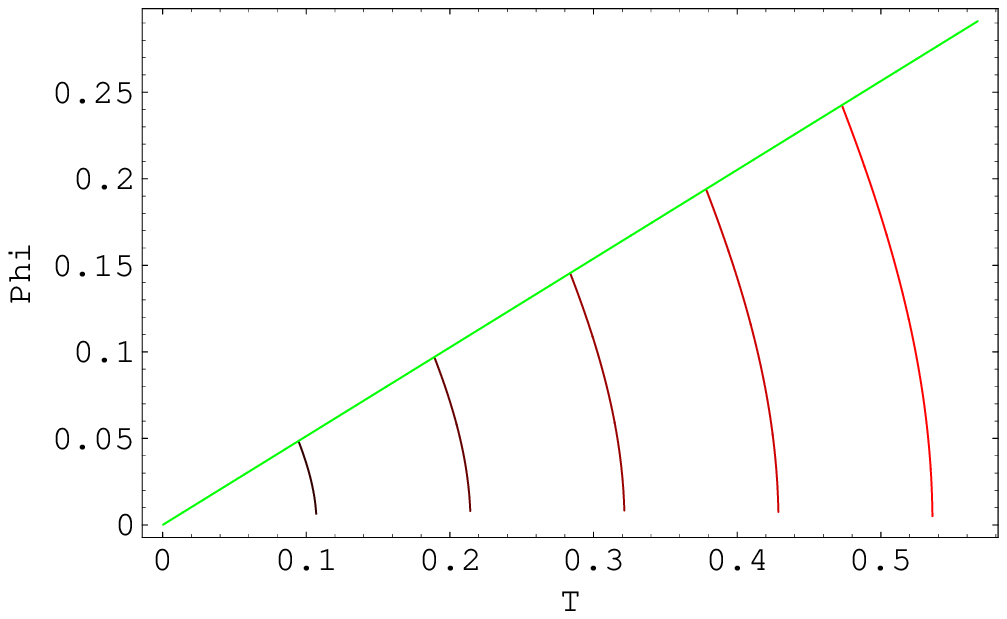, width=.4\textwidth}{$r_0-q$ phase diagram of
7-dimensional R-charged black hole with $q_1=q, q_2=0$.}{$T-\phi$
phase diagram of 7-dimensional R-charged black hole with $q_1=q,
q_2=0$.}

\DOUBLEFIGURE[t]{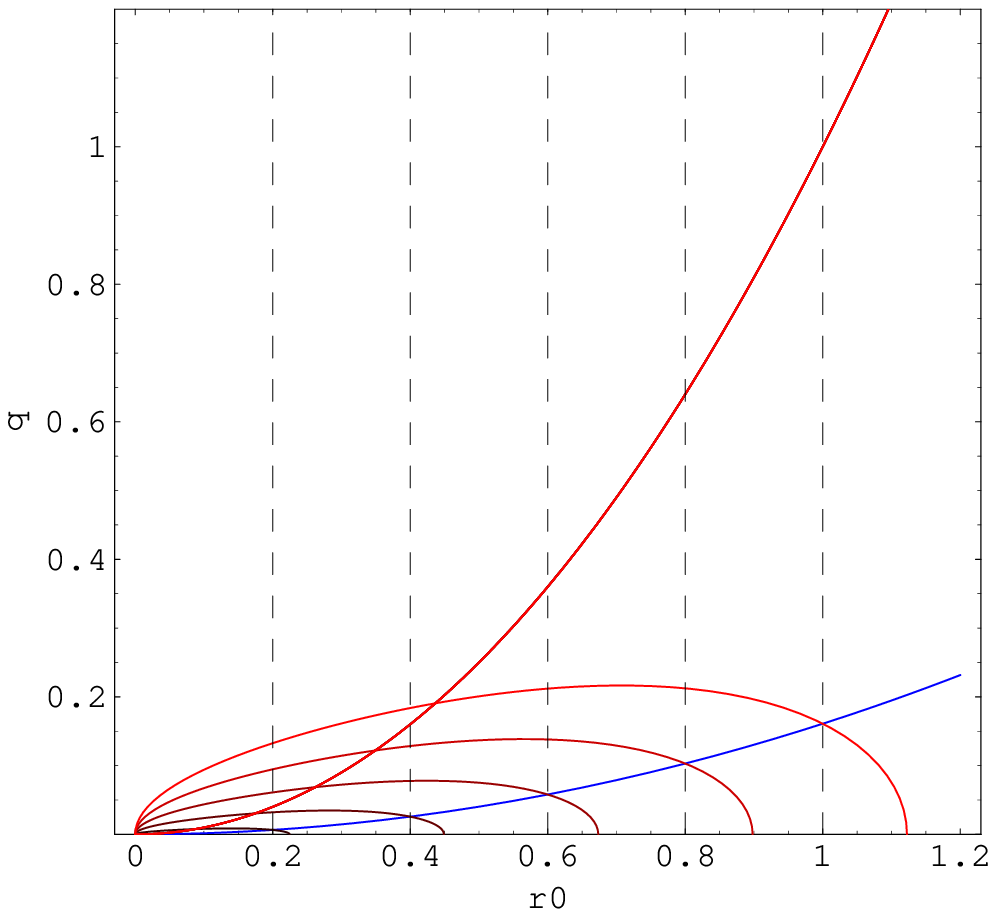, width=.4\textwidth}
{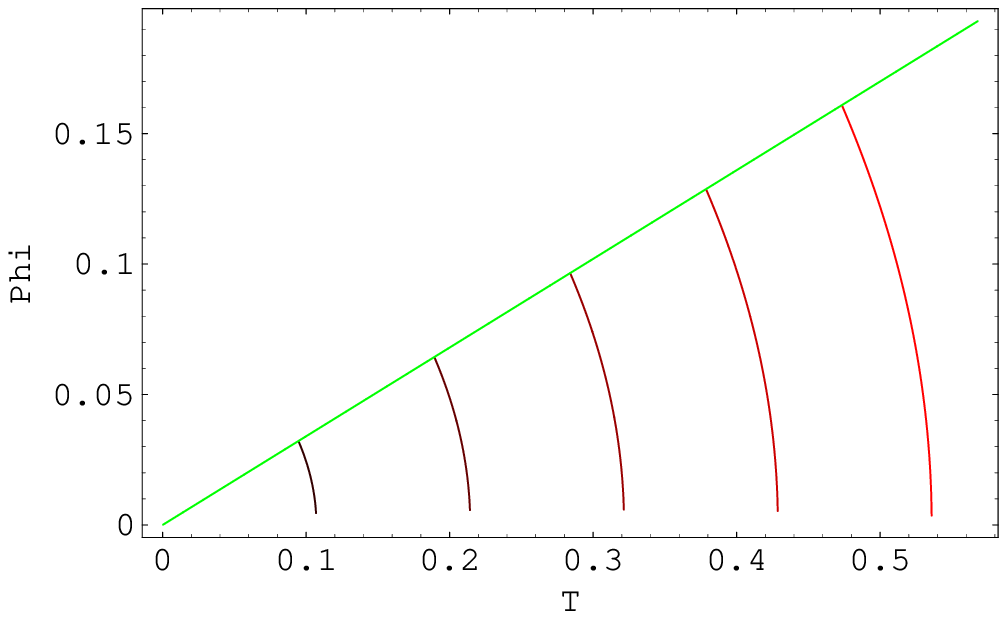, width=.4\textwidth}{$r_0-q$ phase diagram of
7-dimensional R-charged black hole with $q_1= q_2=q$.}{ $T-\phi$
phase diagram of 7-dimensional R-charged black hole with $q_1=
q_2=q$.}

We consider the case $r_0 >r_{IR}$. Because $\mu l^2=r_0^6
{\mathcal{H}}_{1}(r_0){\mathcal{H}}_{2}(r_0)/4$ approaches $r_0^6/4$
when $r_0$ goes to infinity, $\frac{1}{2}r_{0}^6-\mu l^2>0$ can be
easily satisfied for some $r_0$ big enough. Thus as argued in five
dimensional case, for any values of $q_i$'s, there always exists a
value of $r_{0}$ which is denoted by $r_{0c}(q_i)$ such that
$\frac{1}{2}r_{0}^6-\mu l^2>0$ if $r_0>r_{0c}(q_i)$. Therefore, we
can always find an IR cutoff $r_{0c}(q_i)<r_{IR}<r_0$ which
satisfies $\frac{1}{2}r_{IR}^6-\mu l^2>0$.  This means that
introducing a proper IR cutoff can lead to a confinement phase. The
deconfinement transition happens when the action difference
(\ref{4eq81}) changes its sign.

In figure 15 and 17 we plot the $r_0-q$ phase diagrams for the
case of $q_1=q,q_2=0$ and $q_1=q_2=q$, respectively. The five
solid curves correspond to the phase transition curves, and each
curve has a fixed IR cutoff $r_{IR}$. With the color changing from
black to red, the values of $r_{IR}$ increase from $0.2$ to $1.0$
with a step $0.2$. The dash curves represent
$r_0=0.2,~0.4,~0.6,~0.8$ and $1.0$, respectively. The blue curves
stand for the requirement of $r_0>r_{IR}$. In these figures, the
red curves starting from the origin correspond to $q^2=3r_0^4$ and
$q^2=r_0^4$, respectively. They are local thermodynamic stability
curves, determined by the Hessian of the Euclidean action with
respect to $r_0$ and $q_i$ with $\beta$ and $\Phi_i$ fixed. Thus
only the regions below these blue curves satisfy the condition
$r_0>r_{IR}$. As a result, The thermodynamics is always local
stable in those regions.

Figures 16 and 18 give the $T-\phi$ phase diagrams for the case of
$\Phi_1= \phi,~\Phi_2=0$ and $\Phi_1=\Phi_2=\phi$, respectively. The
green curves correspond to the requirement $r_{IR}<r_0$. With the
colors changing from black to blue, the value of $r_{IR}$ increases
from $0.2$ to $1.0$ with a step $0.2$. Again we only give the
regions where the deconfinement transitions happen.

\section{Conclusion}
In this paper we have studied in grand canonical ensemble the
Hawking-Page phase transition associated with decoupling limits of
black Dp-branes ($0\le p\le 4$) and R-charged $AdS_5$, $AdS_4$ and
$AdS_7$ black holes coming from spherical reduction of rotating
black D3-, M2- and M5-branes respectively. The Hawking-Page phase
transition can be identified with the confinement-deconfinement
phase transition of dual SYM theories at finite temperature.

 For the case of the near horizon geometries of black Dp-branes, there does not exist any phase transition
 for the dual SYM theories in non-compact spacetime $S^1\times R^p$,
 although when $p\ne 3$, the dual theories are not conformal. The Euclidean action
difference between the near horizon geometries of  black Dp-branes
and BPS Dp-branes are always negative, which means that the dual
field theories are always in the deconfinement phase. When we
introduce an IR cutoff, as the case of hard-wall AdS/QCD model, a
confinement phase can be realized. And then the deconfinement
transition for the dual SYM theories occurs at some critical
temperature which is determined by the IR cutoff.

The Hawking-Page phase transition also does not appear for the
R-charged $AdS_5$, $AdS_4$, and $AdS_7$ black holes with Ricci flat
horizon. These black holes are dual to some R-charged supersymmetric
field theories on the AdS boundary. When we introduce a proper IR
cutoff, once again, we can realize the deconfinement phase
transitions for those field theories. We have analyzed in some
detail the phase diagrams associated with those R-charged black
holes.

\acknowledgments
 LMC and YWS would like to thank Bin Hu, Ding Ma and Wei-Shui Xu
for useful discussions and kind help. This work was supported
partially by grants from NFSC, China (No. 10325525 and No.90403029),
and a grant from the Chinese Academy of Sciences.

\end{document}